\documentclass[%
 reprint,
 amsmath,amssymb,
 aps,
prstab,
floatfix,
]{revtex4-2}

\usepackage{graphicx}
\usepackage{dcolumn}
\usepackage{bm}

\usepackage{caption}
\usepackage{subcaption}

\usepackage{url}

\begin{document}

\preprint{APS/123-QED}

\title{The anomalous skin effect and copper cavity operation at cryogenic conditions}

\author{Ulrich Ratzinger}
\email{u.ratzinger@iap.uni-frankfurt.de}
\author{Huifang Wang}
\affiliation{Institute for Applied Physics, Goethe University, Max-von-Laue-Strasse 1, 60438 Frankfurt am Main, Germany \\
Helmholtz Forschungsakademie Hessen für FAIR (HFHF), GSI Helmholtzzentrum für Schwerionenforschung, Campus Frankfurt}

\date{\today}

\begin{abstract}
A “geometric model” based on the assumption of a spherical Fermi-surface and using the equivalent skin layer model allows to calculate the surface resistivity of cold normal conducting cavities by including the anomalous skin effect ASE. This is true for copper, silver, or other “simple” metals at operation temperatures down to 4 K and for operating frequencies up to about 2.5 GHz. An electron loss criterion for leaving the skin layer is applied to individual electron positions in the 6-dim phase space. An exact solution for this conduction electron model in skin layers was derived. It is compared with measurements and with predictions from the "universal" ASE diffusion model as formulated by Pippard, Reuter-Sondheimer, Chambers and others. The geometric model predicts up to 15\% higher losses in the region, where the electron mean free path is about 5 times longer then the classical skin depth. Besides that, both models agree very well in the prediction of the surface resistance. 

The derived formalism is useful to calculate the RF power losses on cavity or waveguide walls at cryogenic temperatures. Examples are accelerator cavities, power-couplers in superconducting cavities, or wake-field losses along cold beam tubes of high-current storage rings. Only the direct current DC conductivity dependence on temperature is to be known for the given material. 

Vice-versa, the RF – conductivity from cavity measurements can be directly converted into corresponding effective DC values: Even if the DC values of the bulk material are known very well, there might be drastic local aberrations from those conductivity values after the full manufacturing process has been finished. Table \ref{tab:tableA} in the appendix allows for an easy conversion in both directions. 

A pulsed ion-linac operation with copper-cavities at cryogenic temperatures around 40 K is discussed. An overall advantage in costs by considerably higher acceleration fields is expected for linear synchrotron injectors or medical linacs.

After describing the geometric ASE model, own cryogenic measurements on a coaxial quarter-wave resonator are shown as well as fits on two published measurements. The geometric model agrees very well to the proven frequency dependence of the surface resistance - changing from 1/2 towards 2/3 in the extreme ASE regime.

H-type cavities are attractive candidates for such a cryogenic ion linac. They will allow a further effective voltage gain increase beyond 10 MV/m. The latter value was demonstrated successfully with an IH-type cavity at room temperature operation. Cooling simulations about the pulsed heat transport from the cavity surface into the bulk copper are showing promising results. 
\end{abstract}

\maketitle


\section{INTRODUCTION}

Normal conducting RF structures might exploit the significantly higher conductivity of simple metals like copper and silver, when operated at temperatures of liquid nitrogen and below. On the other hand, the anomalous skin effect ASE in many cases reduces that gain considerably\cite{pippard1954anomalous, reuter1948theory, chambers1952anomalous}. Some intense studies and experiments were performed recently, which are quite encouraging with respect to increased field levels at cavity operation temperatures between 30 K and 70 K\cite{jacewicz2020temperature, nordlund2012defect, Vernieri_2023, nasr2021experimental, braun2003frequency, CahillIPAC2016-MOPMW038}. While these studies are motivated by applications in electron beam acceleration at GHz – frequencies, the aim of this paper is to find applications in the 100 MHz to 2.5 GHz range, typical for proton and ion acceleration. It is assumed, that the improvement in maximum surface field levels will be similar to what was demonstrated already for electron accelerator cavities\cite{thesis2023wang, wangipac2023-tupa189}. This should allow to find a good compromise between reduced RF power needs and a reduced total linac length to save building costs.

The paper describes a concept and derives formulas to calculate the resistive RF wall losses under cryogenic conditions. There are some hints (see for examples Refs. \cite{caspers1997surface, weingarten}), that the diffusion model is too optimistic in resistivity predictions – at least in the frequency range below 2.5 GHz, where no significant losses by surface contamination and surface roughness are expected\cite{liew2014signal}: It is hereby assumed, that $R_a=0.3$ micrometer surface roughness is reached on all cavity surfaces.

H-Mode structures of the IH- and of the CH-type have rather small cavity diameters and simple drift tubes at the frequency range under discussion\cite{ratzinger2019combined}. Moreover, they can achieve effective acceleration voltage gains above 10 MV/m already at room temperature operation\cite{broere1998high}. With including the new techniques of 3d – printing of stainless steel and copper components\cite{hahnel2023additive} it should be possible to reach attractive copper cavity solutions for operation temperatures around 40 K and with effective voltage gains around 15 MV/m. This is at least a factor 3 above the state of the art in pulsed ion acceleration up to around 150 $A MeV$.

A very important point is the temperature stability of the cavity surface during the RF pulse. Heat transport calculations and temperature profiles across the cavity walls are presented and discussed.

Other topics to apply the geometric model are for example copper components in superconducting cavities – like power couplers. It is of great importance to know exactly the thermal losses at these surfaces, which can’t be cooled efficiently in an easy way\cite{fouaidy2006rrr}, and which are drifting to intermediate temperature levels depending on the operating conditions. The surface resistance of cold beam pipes is as well an important subject, as the anomalous skin effect can cause a large contribution to the beam driven wall losses \cite{caspers1997surface}, and direct cooling is not possible in many cases. In space-flight, many RF devices are exposed to varying temperatures in the range as discussed here. For these applications, it is sufficient to only know the resistive part $R$ of the surface impedance $Z$.

\section{ESTIMATION OF THE ANOMALOUS SKIN EFFECT IN MICROWAVE CAVITY WALLS}

\subsection{Overview}

Models about the ASE were developed from the beginning for a wide range of frequencies and metals. Models based on the so-called diffusion model were developed and experimentally tested by A.B. Pippard and others like G.E. Reuter, E.H. Sondheimer, R.G. Chambers. The theory is valid over a wide range of frequencies.

These descriptions were successful and have been proven experimentally. However, the RF experiments often show reduced performance against the theoretical predictions also at lower frequencies, where the surface roughness should be not a big issue. This gave the motivation to describe the anomalous skin effect up to around 2.5 GHz and at practicable RRR values with another approach, and to be focused on the change in wall resistance only. This is completely sufficient to describe the RF losses in cavities or the attenuation along wave guides. A “geometric ASE model” is developed, where a well-defined starting condition for the 6-D phase space distribution of electrons is exactly solved with respect to the percentage of electrons, which are leaving the skin sheath within their free path length between two collisions at collision time $\tau$. 

The upper frequency limit for this model at around 2.5 GHz is set by the condition $\omega \tau \ll 1$, which is to be fulfilled at all temperatures. At higher frequencies, the influence of the ASE on the one hand is overestimated by this model, as the shorter RF period results in a lower loss probability of current carrying electrons which is not included here. On the other hand, the ASE is underestimated, as small angle scattering of electrons will become important at very small skin depth to free electron path - ratios.

\subsection{Assumptions for the geometric model}
The main parameters of copper are introduced from the beginning to make the assumptions clear by numerical cross-checks. However, the model can be applied to other “simple” metals as well by adapting the specific parameters.
\begin{table}[b]
\caption{\label{tab:table1}
Key parameters for copper.}
\begin{ruledtabular}
\begin{tabular}{lc}
Fermi-energy $E_F/eV$ & 7.0\\
Fermi-velocity $v_F/m/s$ & $1.57\cdot 10^6$\\
Free electron density $n_e/m^{-3}$ & $8.5\cdot 10^{28}$\\
Electron mass $mc^2/keV$ & 511\\
Parameters at 293 K: & \\
Spec. electr. resistivity $\rho_n/\Omega m$ & $1.7\cdot 10^{-8}$\\
El. Conductivity $\sigma_n/S/m$ & $5.9\cdot 10^7$
\end{tabular}
\end{ruledtabular}
\end{table}

From now onwards, temperature dependent parameters contributing to the anomalous skin effect are denoted by an index $a$ like $\sigma_a$, $\rho_a$, $\delta_a$, while parameters at room temperature are denoted by an index $n$. Parameters from the normal skin effect theory are denoted by an index $c$ like $\delta_c$.

From the parameters in Table \ref{tab:table1} the electron collision time $\tau_n$ and the mean free path $l_n$ at 293 K can be derived classically:
\begin{equation}
\sigma_n=\frac{n_e e^2\tau_n}{m}
\end{equation}
This results in
\begin{equation}
    \tau_n=2.46\cdot 10^{-14}s \tag{1a}
\end{equation}
and in a mean free electron path length of
\begin{equation}
    l_n=v_F \cdot \tau_n=3.9\cdot 10^{-8}m=0.039\mu m. \tag{1b}
\end{equation}
It means, that $\tau$ is very short against the RF period $T$. The “normal” skin depth in copper at 293 K and at 340 MHz equals to:
\begin{equation}
    \delta_n=\sqrt{\frac{2}{\omega\cdot \mu_0 \cdot \sigma_n}}=6.56\cdot 10^{-2}\cdot f^{-1/2}=3.56 \mu m.
\end{equation}

\begin{figure}[!h]
\includegraphics[width=\columnwidth]{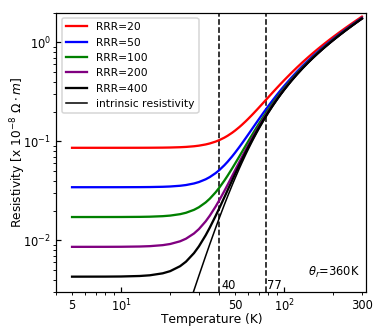}
\caption{\label{fig:fig1} Dependence of the specific electrical resistivity in copper from its RRR - value and from temperature by applying the Bloch-Grüneisen formula, Eq. (\ref{eq-bloch_gruneisen}).}
\end{figure}

No ASE will happen at room temperature operation as $\delta$ is by two orders in magnitude larger than the mean free electron path. However, the free path is increasing proportionally to $\sigma(T)$ and the skin depth is reduced with $\sigma(T)^{-1/2}$. It becomes clear from Fig. \ref{fig:fig1}, that the ASE will become an issue, as soon as the RRR - values of the copper surface are high enough. The temperature dependence of the electric specific resistance is described by the Bloch-Grüneisen formula\cite{matula1979electrical}, which in case of copper reads
\begin{eqnarray}
	\rho(T) =&& \frac{\rho(300K)}{RRR}+4.226\cdot \left( \frac{T}{\theta_r} \right)^5\cdot \rho(\theta_r) \nonumber \\
    &&\cdot \int_0^{\theta_r/T}\frac{t^5}{(e^t-1)(1-e^{-t})}dt \label{eq-bloch_gruneisen}
\end{eqnarray}
where $\rho(\theta_r)$ is the electrical resistivity at the resistive Debye temperature.

A realistic RRR - value for larger quantities of deliverable material is $RRR$=400 for Cu-OFE, trade names are CW009A in Europe, C10100 in USA. For envisaged operation temperatures at and above 40 K, RRR – values around 200 are sufficient (compare Fig. \ref{fig:fig1}). This results in parameter limits for the application of the geometric model on copper:
\begin{eqnarray}
	&&\sigma(4.2K)=1.77\cdot 10^{10} S/m;\quad \tau(4.2K)=4.92\cdot 10^{-12}s; \nonumber  \\ &&l(4.2K)=11.7 \mu m, \quad \omega \tau=0.077 \text{ at 2.5 GHz.} \label{eq:boundary}
\end{eqnarray}
Compared to the corresponding electron collision rate of down to 135 GHz at 4.2 K, one can conclude, that up to around 2.5 GHz, the electron collision rate at low temperature is still high against the RF period.

\begin{figure}[!h]
\includegraphics[width=\columnwidth]{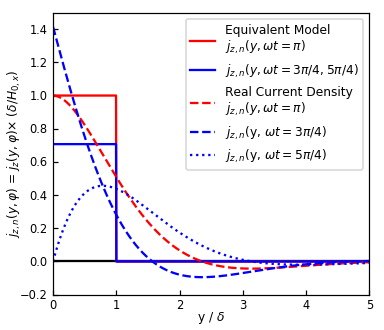}
\caption{\label{fig:fig2}Current density amplitude across the conducting sheath according to the normal skin effect and equivalent model\cite{jackson1999classical} assumption. The red curves correspond to the field maximum, while the blue curves are relevant to the situation ±45° before and after the maximum.}
\end{figure}

The following assumptions are made:
\begin{itemize}
	\item[1.] The cavity surface is locally described as a plane perpendicular to the $y$-axis
	\item[2.] 	The tangential H-field is along the $x$-direction, the corresponding surface RF current and the Poynting-vector along the $z$-direction, the cavity E-field along the $y$-direction. This meets the situation in waveguides and cavities, but is not suited for calculating thin film penetration etc.
	\item[3.] The RF current density is constant across the skin layer and zero beyond (equivalent conductive layer model (see \cite{jackson1999classical} and Fig. \ref{fig:fig2}), resulting in a correct surface resistance in case of the normal skin effect. 
	\item[4.] 	Electrons hitting the surface are assumed \\
	\quad - to show specular reflection ($p=1$) \\
	\quad - to be completely diffusely reflected ($p=0$) \\
    Both cases are compared with measurements to find an according description finally. The model allows for readjustments, as $p$ of cause also depends on the surface quality.
	\item[5.] A Fermi-Dirac distribution at 0 K for free electrons is assumed. This sphere is displaced by $\Delta v_F$ when electric surface current is transported. Current carrying electrons keep their velocity along the free path $l$. This defines the upper frequency limit for this model by the condition $\omega \tau << 1$
	\item[6.] Electrons stop contributing to the RF current as soon as they leave the skin layer along their free path between two collisions
	\item[7.] Anomalous electron losses within a given velocity space will be partly compensated by electrons accelerated from the core into the available states
\end{itemize}

Assumption 3. is visualized by Fig. \ref{fig:fig2}. The current-density distribution across the conducting metal sheath (Ref. \cite{jackson1999classical}) is plotted at $\varphi=0, -\pi/4$ (corresponding to the maximum surface current), and $-\pi/2$. The according $\theta$-functions describing the equivalent skin model at those phases are plotted as well.

Current density:
\begin{subequations}
\begin{equation}
    j_z(y,t)=\frac{-1+i}{\delta}\cdot H_0 \cdot e^{-y/\delta} \cdot e^{i(y/\delta-\omega t)} \label{eq:eq_complex_current}
\end{equation}
The real part is
\begin{eqnarray}
	&&j_z (y, \varphi)=\sqrt{2}\cdot H_{0,x}/\delta \cdot e^{-y/\delta}\cdot cos(y/\delta+\varphi);   \nonumber \\
 && \varphi =3\pi/4-\omega t; \quad y\ge 0. \label{eq-j_1}
\end{eqnarray}
Integration over $y$ results in the surface current\\
$\kappa_z(t)=-H_x(t)=-H_{0,x}\cdot cos(\omega t)$. \\
Equivalent current-density model:
\begin{equation}
	j_z(y,t)=-H_{0,x}/\delta \cdot cos(\omega t);\quad 0\le y \le \delta \label{eq-j_equiv}
\end{equation}
\end{subequations}
and elsewhere \quad $j_z(y,t)=0$. \quad $H_{0,x}\equiv$ Tangential magnetic surface field amplitude. \\

\begin{figure}[!h]
\includegraphics[width=\columnwidth]{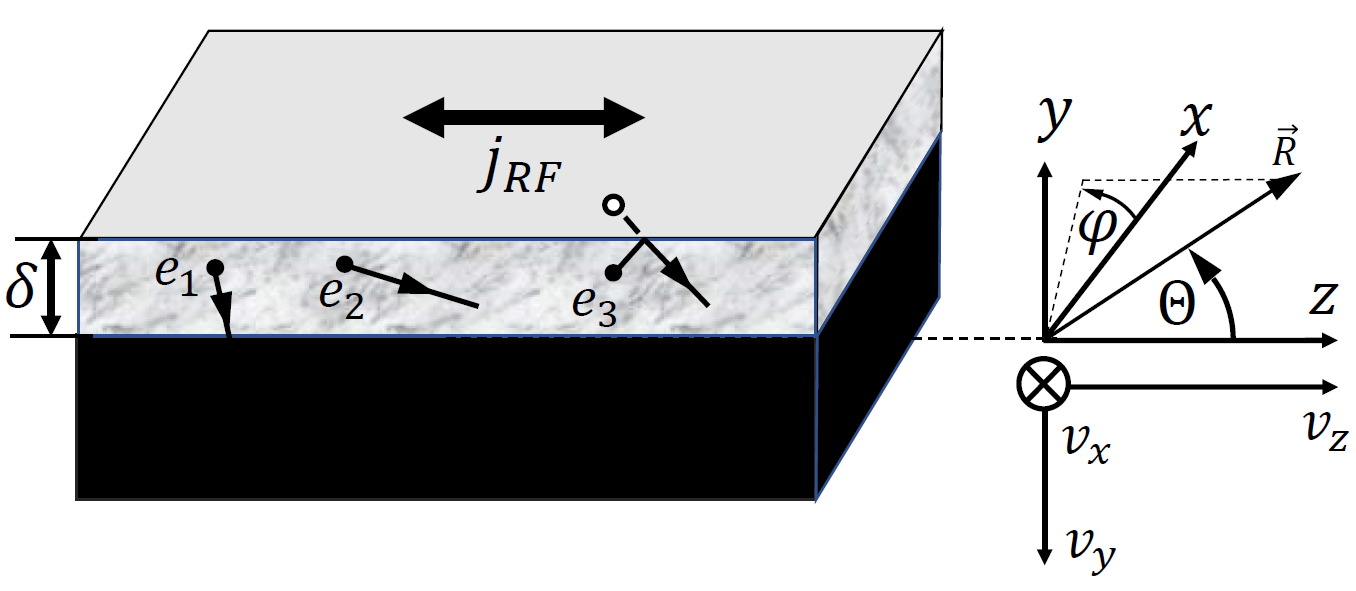}
\caption{\label{fig:fig3}Orientation of the skin layer, the RF current and the coordinate systems to calculate the anomalous skin effect. The metal surface corresponds to $y=\delta$. Additionally, paths of electrons are shown, $e_3$ is an example for specular reflection. For convenience, $v_y$ is oriented into the metal.}
\end{figure}

It should be noted that assumption 3. excludes a calculation of the complex surface impedance, which is defined for good conductors in coordinates as defined above as
\begin{subequations}
    \begin{equation}
        Z=E_z(y=0)/H_{0,x}=R+iX
    \end{equation}
The reactance can be written as
\begin{equation}
    X=i\omega L,
\end{equation}
\end{subequations}
$L$ denoting the conductor inner field inductance per unit area. This reactance has the same magnitude as the resistance (see Eq. (\ref{eq:eq_complex_current})) and can be usually neglected against the cavity inductance or against the wave guide inductance per unit length for applications as discussed in this paper. Details on $X$ with respect to the ASE are given for instance in Ref. \cite{reuter1948theory}. However, assumption 3. should be well suited to calculate the surface resistance - like in case of the normal skin effect. Figure \ref{fig:fig2} gives an idea, how much the current density profile is changed during one RF half period, and how the equivalent model compares to it.
(Coordinates like in Fig. \ref{fig:fig3}, but the $y$-axis in Eqs. (\ref{eq-j_1}, \ref{eq-j_equiv}) is oriented into the metal, and $y$=0 corresponds to the metal surface.)

With respect to assumption 4., the graphs in Refs. \cite{reuter1948theory, chambers1952anomalous} from results of the ASE diffusion model show up to about 25\% difference in surface resistance with respect to a variation of $p$. In this paper, like in Ref. \cite{reuter1948theory}, a transition from lower to higher $p$-values with decreasing temperature is suggested and justified with a decreasing averaged reflection angle. Best fits to measurements are resulting from that concept. Especially at higher frequency, the technical limits in surface treatment are of cause contributing additionally. The 0K distribution function $f_0(v)$ (Assumption 5.) is well justified in case of the ASE description, as the thermal energy is three orders smaller in magnitude than the Fermi energy. The diffusion ASE model makes a very similar approach by setting the derivative $f_0'(v)$ equal to the delta function $\delta(v-v_F)$.

A consequence from assumptions 5. and 6. is, that an exponential decay of the current contribution along each single electron path is not included here. This agrees to literature, where the electric resistance at lower temperatures is mainly attributed to Umklapp-scattering processes with phonons\cite{ashcroft1976solid}, which change the electron path direction significantly in one collision. The mean free path concept and the insertion into the free electron equation of motion by an additional term (see Eq. (\ref{eq:eqASE})) to add the anomalous losses, is justified by that fact. Only at very low temperatures of a few Kelvin, Umklapp-processes become very rare and small – angle scattering is assumed to dominate the electron-phonon interaction. This might be the reason, that the geometric model predicts a smaller surface resistance then the diffusion model at very low temperatures and very high RRR - values (see Fig. \ref{fig:fig11}, top).

The Fermi-energy ($7 eV$) is about three orders higher then the thermal energy ($meV$) while the shift of the Fermi-surface by typically operated cavity levels is again three orders in magnitude smaller ($\mu eV$) then the thermal energy. Statistically, the localization of the electrons in phase space contributing to the surface currents are well described by the geometric model. The authors of the diffusion model already demonstrated, that details of the individual Fermi-surfaces of most common metallic conductors are not needed to describe quantitatively the anomalous skin effect in cavities and wave guides. Instead, the free electron model is sufficient. This applies even more, if the surface resistance is investigated only.

The quality factor of a cavity with volume $V$ and surface $A$ is proportional to $\delta^{-1}$ at a given geometry\cite{jackson1999classical}:
\begin{equation}
	Q=g\cdot \frac{V}{A\cdot \delta} \label{eq:eqQVgA}
\end{equation}
The geometric factor $g$ is due to the individual cavity geometry. From a measured $Q(T)$ at a temperature T, one can derive $\delta(T)$, and by that parameter gets access to the conductivity. However, the RF conductivity is reduced against its DC-value as soon as the anomalous skin effect is contributing and the skin depth will increase according to Eq. (\ref{eq:eqQVgA}).

This conductivity reduction at each temperature will be derived now by applying the geometric model. 

At an operation temperature T, the DC-conductivity reads
\begin{subequations}
	\label{eq-electron_motion}
	\begin{equation}
		\sigma(T)=\frac{n_e e^2 \tau(T)}{m}. \label{eq:eq4a}
	\end{equation}
\text{The free conducting electron motion along $z$ at AC fre-}
\text{quency $\omega$ is described by:}
\begin{equation}
	m\dot{v}_z+m\frac{v_z}{\tau}=-e\cdot E_z, \label{subeq:eq4b}
\end{equation}
\begin{equation}
	m\cdot \left( \frac{1}{\tau}+i\omega \right)v_z = e \cdot E_z \label{subeq:eq4bb}
\end{equation}
\begin{equation}
	j_z=-n_e\cdot e \cdot v_z=\sigma \cdot E_z.   \label{subeq:eq4c}
\end{equation}
\text{This results in} 
\begin{eqnarray}
	&&\sigma = \frac{n_e\cdot e^2\cdot \tau\cdot (1-i\omega \tau)}{m\cdot(1+\omega^2\tau^2)}\to \frac{n_e\cdot e^2\cdot \tau}{m} \nonumber \\ 
 &&\text{for} \quad f\le2.5 GHz, \text{see Eq. (\ref{eq:boundary})}  \label{subeq:eq4cc} 
\end{eqnarray}
\text{resulting in the simplified differential equation in equili-}
\text{brium}
\begin{equation}
    m\frac{v_z}{\tau}=-e\cdot E_z \label{subeq:eq4f}
\end{equation}
\end{subequations}
The imaginary term in Eq. (\ref{subeq:eq4bb}) can be neglected for the range of RRR - values and frequencies as specified before.  $v_z$ in Eq. (\ref{subeq:eq4c}) is illustrated in Fig. \ref{fig:fig4} and describes the shift $\Delta v_F$ of the Fermi-sphere in an electric field.

However, the RF surface conductivity $\sigma_a(T)$ might be reduced, with consequences for the skin depth:
	\begin{equation}
		\delta_a(T)=\sqrt{\frac{2}{\omega\cdot \mu_0 \cdot \sigma_a(T)}}. \label{eq:eq5a}
	\end{equation}
The reason are electrons with a velocity component $v_y$, large enough to leave the skin layer between two collisions. By that they enter a field and current free zone.

It is useful to define the following parameters:
\begin{subequations}
	\begin{eqnarray}
		h_l&&=\frac{\delta_a(T)}{l(T)}=\sqrt{\frac{2}{\omega\cdot \mu_0 \cdot \sigma_a(T)}}\cdot 
		\frac{n_e\cdot e^2}{v_F\cdot \sigma(T)\cdot m_e} \nonumber \\
		&&=\frac{(\sigma_n)^{3/2}}{\sigma(T)\cdot \left( \sigma_a(T)\right)^{1/2}}\cdot \frac{1.7\cdot 10^6}{f^{1/2}}.  \label{eq:eq5b}
	\end{eqnarray}
	\begin{equation}
		h_0=\frac{\delta_c(T)}{l(T)}=\frac{(\sigma_n)^{3/2}}{(\sigma(T))^{3/2}}\cdot \frac{1.7\cdot 10^6}{f^{1/2}} \label{eq-define_h0}
	\end{equation}
	\begin{equation}
		q_0=\frac{(\sigma_n)^{3/2}}{(\sigma_a(T))^{3/2}}\cdot \frac{1.7\cdot 10^6}{f^{1/2}} \label{eq-define_q0}
	\end{equation}
\end{subequations}
$\delta_c(T)$ denotes the classical skin depth.

\subsection{Conducting electrons in velocity space}
\begin{figure}[!h]
	\centering
	\includegraphics[width=\columnwidth]{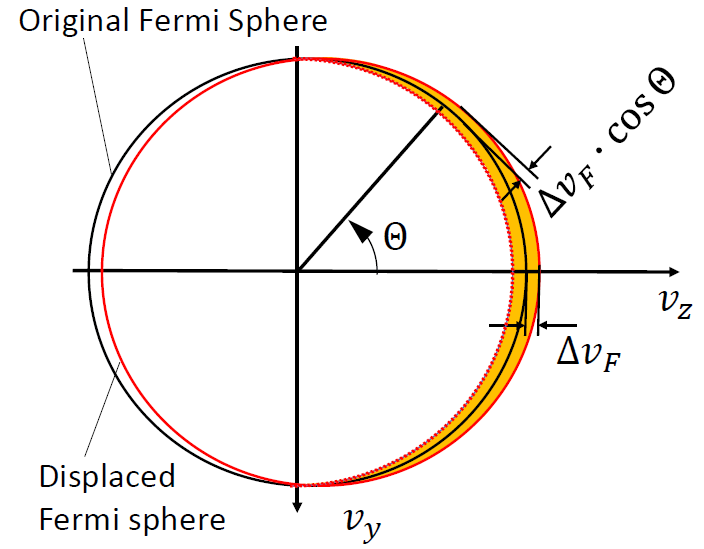}
	\caption{\label{fig:fig4} Conducting sheath (in orange) with maximum thickness $2\Delta v_F$ in $v_z$–direction.}
\end{figure}

When RF current is flowing, the current carrying electrons in velocity space are those within the hemispherical shell as sketched in Fig. \ref{fig:fig4} in orange. At first, the integral in momentum space over this shell is carried out. $\Delta v_F$ in Fig. \ref{fig:fig4} corresponds to $v_z$ in Eqs. (\ref{subeq:eq4b}, \ref{subeq:eq4bb}) and is typically at least 6 orders in magnitude smaller than $v_F$.

The relative current density contribution in local space from a solid angle element $d\Omega$ at $\Theta$ in electron velocity space varies then (see Fig. \ref{fig:fig4}) with
\begin{equation}
	d n_c(\Omega) \sim 2\Delta v_F \cdot v_F^2 \cdot cos\Theta \cdot d\Omega.
\end{equation}
Each electron in the conducting half shell contributes to the current density $j_z$ proportionally to its velocity
\begin{equation}
    v_z=v_F\cdot cos\Theta \nonumber
\end{equation}

The current density as used in Eq. (\ref{subeq:eq4c}) results then from integration over the velocity space of contributing electrons:
\begin{eqnarray}
	j_{z,tot}&&\sim e \int_{\Theta=0}^{\pi/2} \int_{\varphi=0}^{2\pi}v_z(\Omega)d n_c(\Omega)d\varphi d\Theta \nonumber \\
	&&\sim \frac{4\pi}{3}\cdot e\cdot \Delta v_F\cdot v_F^3.\label{eq:eq8}
\end{eqnarray}
By multiplication of Eq. (\ref{eq:eq8}) with the conduction electron density $n_e$ and division by the homogeneously filled volume of the Fermi sphere one gets the expected result:
\begin{equation}
    j_{z,tot} = n_e e \Delta v_F. \label{eq:eq8b}
\end{equation}

For mathematical reasons, this integration is projected into the $v_x v_y$–plane: There, the conducting sheath $\Delta v_z$ has a constant thickness $2\Delta v_F$. However, the $cos\Theta$ - dependence of $v_z$ has to be adapted.

The integration is now performed along the coordinates $u=v_x/v_F$, $w=v_y/v_F$, for the circular area. The integral reads then
\begin{equation}
	j_{z,tot}\sim e\cdot 2\Delta v_F\cdot v_F^3\cdot \int_{-1}^{1}\int_{-(1-w^2)^{1/2}}^{(1-w^2)^{1/2}}(1-u^2-w^2)^{1/2}dwdu. \label{eq:eq9}
\end{equation}

\subsection{Fraction of lost RF conduction electrons in a y-plane within the skin sheath}
\begin{figure}[!h]
	\centering
	\includegraphics[width=\columnwidth]{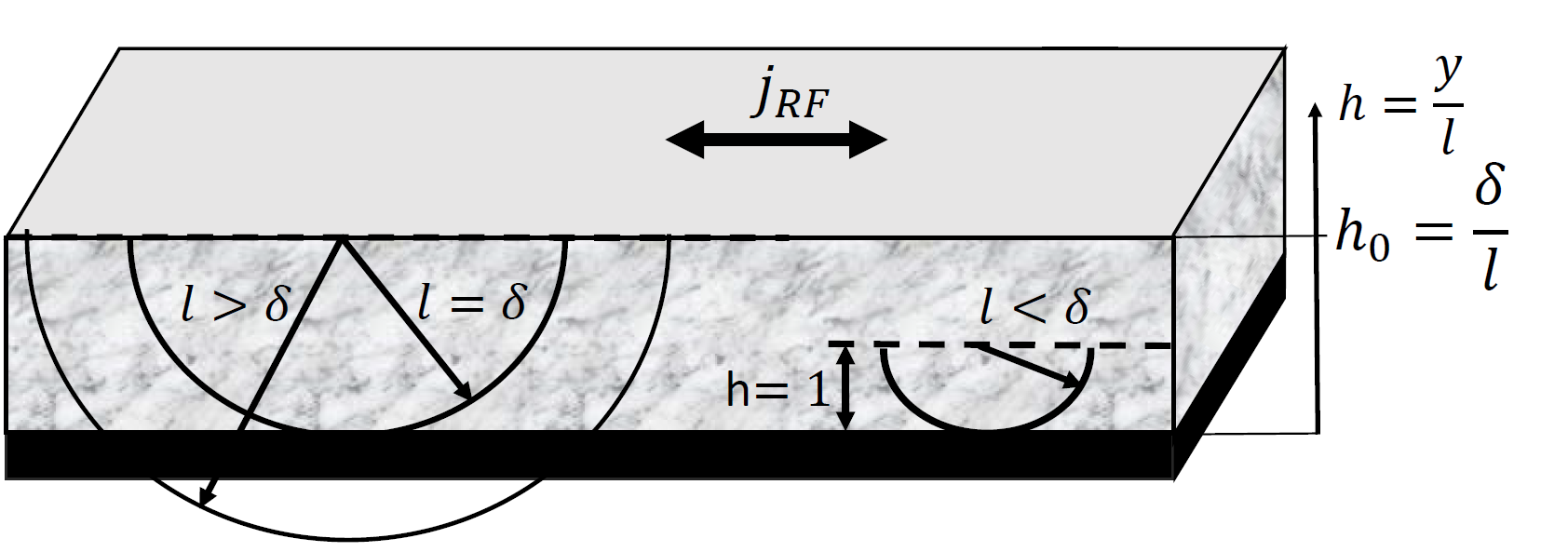}
	\caption{\label{fig:fig5} Illustration of different electron free path length cases and of a reduced loss region up to $h<1$ only for $l<\delta$.}
\end{figure}

In local space the position $y=0$ marks the bottom of the skin layer (see Figs. \ref{fig:fig3} und \ref{fig:fig5}). At a positive position $y$ within the skin layer, an electron will be “lost” as soon as it has a velocity component $v_y$ high enough to pass the projected length $y$ within the collision time $\tau(T)$. Alternatively, an electron at position $y$ will be lost after specular reflection at the metal surface as soon as it has a negative starting component $v_y$, high enough to transit a projected path length $\Delta y=2\delta-y$ within $\tau$. The same result is received, when these electrons are started at $y^{\prime}=2\delta-y$ (see Fig. \ref{fig:fig3}) with positive velocity $v_y$. (Remember: Positive $v_y$ are directed antiparallel to the $y$–axis).

In case of completely diffuse reflection, electrons are lost for RF conductivity in the same way as soon as they reach the upper or lower boundary of the skin sheath. The loss contribution from the positive $v_y$-velocity space is then identical with the negative one and the first one will be just multiplied by two to get the full result.

The loss condition for an RF current carrying electron at $y$ is then
\begin{equation}
	y<v_y\cdot \tau(T), \label{eq:eq10a}
\end{equation}
$y$ will run from 0 to $\delta$ for diffuse reflection and up to $2\delta$ for specular reflection. In the normalized velocity space the integration limits are $0\leq w \leq 1$, $-(1-w^2)^{1/2}\leq u\leq (1-w^2)^{1/2}$.
With
\begin{equation}
	v_y = v_F\cdot sin\Theta \cdot sin\varphi=v_F\cdot w   \label{eq:eq10b}
\end{equation}
the loss condition finally reads
\begin{subequations}
	\begin{equation}
		y<v_F\cdot w\cdot \tau(T)  
	\end{equation}
	\begin{equation}
		w>h(T,y) \quad \text{with}  
	\end{equation}
\end{subequations}
\begin{equation}
	h(T, y)\equiv h=\frac{y}{v_F\cdot \tau(T)} =\frac{y}{l(T)} \label{eq:eqTy}
\end{equation}

With Eq. (\ref{eq:eq4a}) one gets
\begin{equation}
	h=\frac{y\cdot n_e \cdot e^2\cdot c^2}{v_F\cdot \sigma \cdot mc^2}.
\end{equation}
For copper with $v_F$, $n_e$ as given above it results in
\begin{equation}
	h=1.53\cdot 10^{15}\cdot \frac{y}{\sigma}. \label{eq:eq13}
\end{equation}
Finally, with Eqs. (\ref{eq:eq10a}) and (\ref{eq:eq13}) the condition for anomalous losses in copper at depth $y$ is
\begin{eqnarray}
	w>1.53\cdot 10^{15}\cdot \frac{y}{\sigma};\quad &&0<y<\delta \text{ for diffuse reflection}, \nonumber \\
 &&0<y<2\delta \text{ for specular refl.}
\end{eqnarray}

\begin{figure}[!h]
	\centering
	\includegraphics[width=\columnwidth]{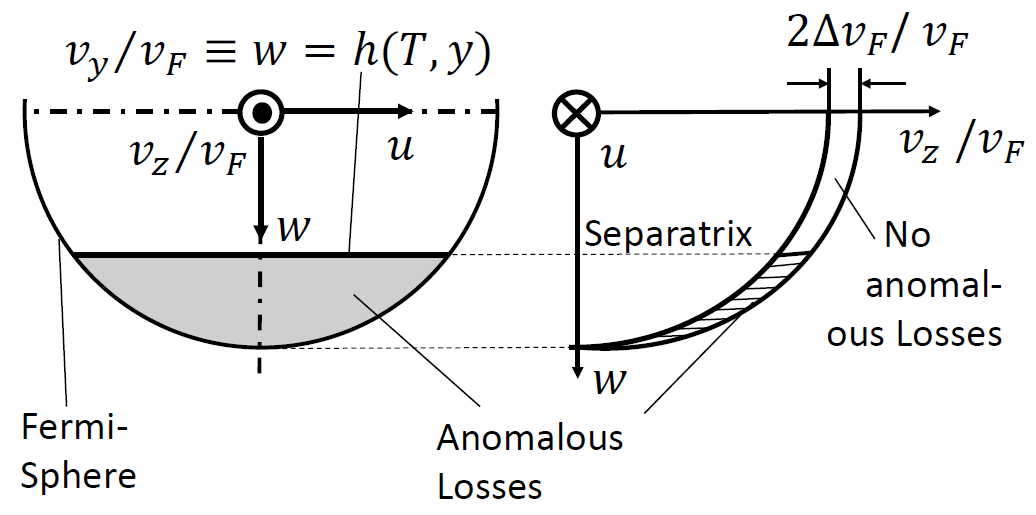}
	\caption{\label{fig:fig6} Illustration of the separatrix between fully conducting and reduced conducting electrons in velocity space at a position $y$ and $h\equiv y/l$, respectively. View on the lower half of the Fermi surface (left) and cut through one quadrant of the Fermi-Sphere (right).}
\end{figure}

As a first task, the percentage of reduced RF conductivity in $z$–direction at a given $h$ – that means in a given $y$–plane $0<y<\delta \quad (2\delta)$ – has to be calculated. The conductivity is reduced by the anomalous skin effect as soon as $h(T, y)<1$. This is illustrated by Fig. \ref{fig:fig6}.

Within a given loss angle around $\Theta$, $\varphi$, the loss time is given by (see Eqs. (\ref{eq:eq10b}, \ref{eq:eqTy} and $\tau=l/v_F$))
\begin{equation}
	\Delta t = \frac{y}{v_y}=\frac{y}{v_F\cdot sin\Theta \sin\varphi}=\frac{h\cdot \tau}{sin\Theta sin\varphi}=\frac{h\cdot \tau}{w}.
\end{equation}
Including this loss mechanism in Eq. (\ref{subeq:eq4b}) as a second "friction term" and by that fulfilling model assumptions 6 and 7, one gets 
\begin{equation}
	m\dot{v_z}+m\frac{v_z}{\tau}+m\frac{v_z}{\Delta t}=-e\cdot E_z \label{eq:eqASE}
\end{equation}
and in equilibrium (same approximation as in Eq. (\ref{subeq:eq4cc} and \ref{subeq:eq4f}))
\[ m\frac{v_z}{\tau}+m\frac{v_z}{\Delta t}=-e\cdot E_z \]
\begin{equation}
	m\cdot \frac{v_z}{\tau^{\prime}}=-e\cdot E_z
\end{equation}
with
\begin{equation}
	\tau^{\prime}=\frac{\tau \cdot \Delta t}{\tau+\Delta t}=\tau\frac{h}{w+h}
\end{equation}
This results in a reduced local conducting electron sheath thickness in velocity phase space:
\begin{equation}
	2\Delta v_z^{\prime}(h, w)=2\Delta v_F \cdot \frac{\tau^{\prime}}{\tau}=2\Delta v_F \frac{h}{w+h},  \text{if} \quad h(T,y)<1
\end{equation}
with $w=v_y/v_F$ as introduced before.

The resulting conductivity at the layer $h=y/l$ within the skin sheath is calculated now by repeating the integration in velocity space like in Eq. (\ref{eq:eq9}), but with replacing the constant $\Delta v_F$ by $\Delta v_z^{\prime}(w)$ within the range $h\le w \le 1, h<1$, and adding the loss free part in the velocity range $0\le w <h, h<1$ (see appendix \ref{app:density_y_plane}).

\subsection{Reduced RF conductivity within the skin sheath} \label{chap:chap2E}

The two assumptions “specular reflection with $p=1$” and “diffuse reflection with $p=0$” are considered separately at first. The integration over $h$ is applied on Eqs. (\ref{eq-j_z1}, \ref{eq-j_z2} and \ref{eq-j_z3}) of appendix \ref{app:density_y_plane}.

In the specular assumption, the integration is from $h=0$ up to $2h_0=2\delta_c/l$ for including the velocity space with negative starting values for $v_y$.

At completely diffuse reflection, as mentioned before the conductivity calculated for positive $v_y$ is just multiplied by a factor two, the $h$-integration is from 0 to $h_0=\delta_c/l$ to get the correct conductivity value. These calculations are shown in appendix \ref{app:density_whole_sheat}.

The anomalous conductivity at that stage is given by (see Eq. (\ref{eq:eq8}) for $j_{z,tot}$)
\begin{equation}
	\sigma_{a,1}=\sigma \cdot\frac{j_{z,a1}}{j_{z,tot}}=\sigma\cdot F(h_0);
\end{equation}
and has to be chosen accordingly for the 4 different cases for $F$ as defined in appendix \ref{app:density_whole_sheat}:\\
\begin{itemize}
	\item[a)] Specular reflection, $p=1$,\quad $2h_0<1$
	\item[b)] Specular reflection, $p=1$,\quad $2h_0\ge 1$  
	\item[c)] Diffuse reflection, $p=0$,\quad $h_0<1$ 
	\item[d)] Diffuse reflection, $p=0$,\quad $h_0\ge 1$.	
\end{itemize}

If a measured DC function $\sigma(T)$ exists, one can apply the geometric model by calculating an $h_0$ with Eq. (\ref{eq-define_h0}). That gives for copper
\begin{equation}
    h_0=\frac{\delta_c}{l}=\left(\frac{\sigma_n}{\sigma(T)}\right)^{3/2}\cdot \frac{1.7\cdot 10^6}{f^{1/2}}
\end{equation}
With $h_0$ inserted in functions $F_{s1,2}$ or $F_{d1,2}$, in the following denoted by $F$, one gets $\sigma_{a,1}$. However, due to reduced RF conductivity the skin depth has to become larger, resulting in an $h_1>h_0$. This leads to a recursion formula with $\sigma \equiv \sigma(T)$:
\begin{subequations}
	\label{eq-recursion_process}
\begin{eqnarray}
	&&\sigma_{a,0}=\sigma;\quad \sigma_{a,1}=\sigma\cdot F(h_0); \quad \sigma_{a,2}=\sigma\cdot F(h_1);...\nonumber \\
 && \text{with} \quad F(h_{-1})=1
\end{eqnarray}
\text{The relation between neighbor-elements of the sequence}
\text{is}
\begin{equation}
	\sigma_{a,i}/\sigma_{a,i-1}=\frac{F(h_{i-1})}{F(h_{i-2})}; \quad i\geq1. \label{eq:eq26}
\end{equation}
\text{For the $h_i$ one gets from the starting condition $\sigma_{a,0}=\sigma$,}
\text{and from Eq. (\ref{eq:eq5a})}
\begin{equation}
	h_{i}=h_{i-1}\cdot \left( \frac{\sigma_{a,i-1}}{\sigma_{a,i}} \right)^{1/2}; \quad i\geq 1. \label{eq:eq28}
\end{equation}
\text{Insertion of Eq. (\ref{eq:eq26}) gives the recursion formula for $h$:}
	\begin{equation} 
		h_{i}=h_{i-1}\cdot \left( \frac{F(h_{i-2})}{F(h_{i-1})} \right)^{1/2}; \quad i\geq 1,
	\end{equation}
\end{subequations}
$F(h_i)$ is calculated by insertion of $h_i$  into the according F-function (\ref{eq:eq23b}, \ref{eq-fs2}, \ref{eq-fd1}, \ref{eq-fd2}), respectively.

\begin{figure}[!h]
	\centering
	\includegraphics[width=\columnwidth]{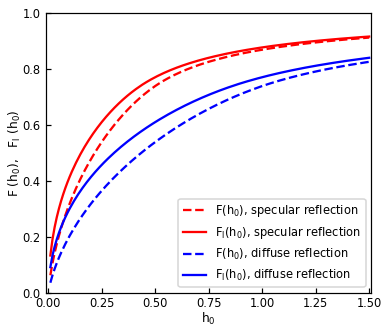}
	\caption{\label{fig-f_function_original} Plot of $F(h_0)$- and $F_l(h_0)$-functions for specular and diffuse reflection.}
\end{figure}

The sequence will converge to an $h_l$ for every starting value $h_0$, with $h_0>0$. With $F(h_l)$ one finally gets the anomalous conductivity
\begin{equation}
	\sigma_a(T)=\sigma(T) \cdot F_l(h_0) \label{eq:eqsigmaa}
\end{equation}
with $h_l=\lim\limits_{i \to \infty}h_i, \quad F_l(h_0)\equiv F(h_l)$ for a given starting value $h_0$. The starting functions $F(h_0)$ and the final functions $F_l(h_0)$ are shown in Fig. \ref{fig-f_function_original} for the most relevant $h_0$-range.

One should also find the DC-conductivity $\sigma(T)$ of an unknown copper quality in case of RF measurements providing $\sigma_a(T)$: As mentioned in many publications (see for example Refs. \cite{pippard1954anomalous, reuter1948theory, chambers1952anomalous}), the RRR - value is very sensitive to manufacturing steps and annealing processes. Therefore, one needs to develop also the reversed calculation process, as described in the following. One might call this an effective DC-conductivity of the cavity surface.

$h$ is now called $q$ for a better distinction. The starting value $q_0$ for copper in that case is given by Eq. (\ref{eq-define_q0})
\begin{equation}
    q_0=\left(\frac{\sigma_n}{\sigma_a(T)}\right)^{3/2}\cdot \frac{1.7\cdot 10^6}{f^{1/2}}
\end{equation}

The ratio between Q - values at temperature T and at room temperature allows directly to calculate $\sigma_a(T)$. Comparison with Eq. (\ref{eq:eq5b}) shows that we have to exchange a factor $\sigma (T)$ in the denominator against $\sigma_a(T)$ by the recursion process. This results in the following recursion process:
\begin{subequations}
	\label{eq-reverse_recursion}
\begin{eqnarray}
	&&\sigma_0=\sigma_a(T); \quad \sigma_1=\frac{\sigma_0}{F(q_0)}; \quad \sigma_2=\frac{\sigma_0}{F(q_1)};...\nonumber \\
 &&\text{with} \quad F(q_{-1})=1.
\end{eqnarray}
	\begin{equation}
		\sigma_i=\frac{\sigma_0}{F(q_{i-1})}; \frac{\sigma_{i+1}}{\sigma_i}=\frac{F(q_{i-1})}{F(q_i)}; i\geq 1, \label{eq:eq33a}
	\end{equation}
	\begin{equation}
		q_i=q_{i-1}\cdot \frac{\sigma_{i-1}}{\sigma_i}\quad \text{and with Eq. (\ref{eq:eq33a})}
	\end{equation}
	\begin{equation}
		q_i = q_{i-1}\cdot \frac{F(q_{i-1})}{F(q_{i-2})};\quad i \geq 1,
	\end{equation}
\end{subequations}
$F(q_i)$ is calculated by insertion of $q_i$  into the according F-function (\ref{eq:eq23b}, \ref{eq-fs2}, \ref{eq-fd1}, \ref{eq-fd2}) - for cases 1 and 2. The transition between both cases is again at $q_i=1$

The recursion will converge to a finite limit $q_l$ for every starting value $q_0$, with $q_0>0$. With this value and $F(q_l)$ one finally gets the DC-conductivity
	\begin{equation}
		\sigma(T) = \sigma_a(T)/F_l(q_0), \label{eq:eq35a}
	\end{equation}
with $q_l=\lim\limits_{i\to \infty}q_i, \quad F_l(q_0)\equiv F(q_l)$ for a given starting value $q_0$.

In contrast to the upper recursion process, now $q_l$ is always smaller than the starting value $q_0$.

By comparing both recursion formulas, they are coinciding with respect to
\begin{equation}
	\frac{\sigma_a(T)}{\sigma(T)}=F(h_l)=F(q_l); \quad h_l(h_0(T))=q_l(q_0(T)). \label{eq:eqhq}
\end{equation}

From Eqs. (\ref{eq:eq5b}, \ref{eq-define_h0}, \ref{eq-define_q0}) one can calculate $q_0$ after knowing $h_0$ and $h_l$, or $h_0$ after knowing $q_0$ and $q_l=h_l$:
\begin{subequations}
    \begin{equation}
        q_0=h_0\cdot \left( \frac{h_l}{h_0} \right)^{3}
    \end{equation}
    \begin{equation}
        h_0=q_0\cdot \left( \frac{q_l}{q_0} \right)^{3/2}
    \end{equation}
    \text{With Eq. (\ref{eq:eqhq}) one gets additionally}
    \begin{equation}
        \frac{h_0}{q_0}=F_l^{3/2}
    \end{equation}
\end{subequations}

That means, that the “forward recursion” according to Eqs. (\ref{eq-recursion_process}, \ref{eq:eqsigmaa}) gives the complete information for a given metal with known $\sigma(T)$ values.

Most authors argue, that the difference between specular  and diffuse reflection is not of significant importance. Ref. \cite{chambers1952anomalous} compares measured curves to calculations and by that comes to the conclusion, that the diffuse assumption results in better fits to measurements (all in the GHz-range).

\begin{figure}[!h]
	\centering
	\includegraphics[width=\columnwidth]{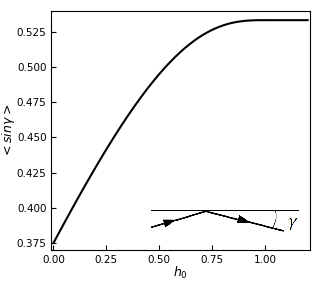}
	\caption{\label{fig:fig9} Plot of the $\langle sin\gamma \rangle$-function.}
\end{figure}

In this paper the dependence of the averaged incidence angle $\gamma$ of reflected electrons, which are contributing to anomalous current losses, is taken into consideration. As stated before, we do not assume a large contribution from surface roughness for the frequency range as discussed here. Each loss is weighted with its current loss contribution. Instead of $\langle \gamma \rangle$, $\langle sin\gamma \rangle = \langle w \rangle$ can be calculated much easier, and reflection is performed at plane $h=0$ with positive $w$ for convenience:
\begin{equation}
	\langle sin \gamma \rangle = \int_{0}^{h_0}\int_{h}^{1}\int_{-(1-w^2)^{1/2}}^{(1-w^2)^{1/2}}w \cdot (1-w^2-u^2)dudwdh /U
\end{equation}
with
\begin{equation}
	U=\int_{0}^{h_0}\int_{h}^{1}\int_{-(1-w^2)^{1/2}}^{(1-w^2)^{1/2}} (1-w^2-u^2)dudwdh
\end{equation}
giving
\begin{eqnarray}
	&&\langle sin\gamma \rangle = (1.5-h_0^2+0.3h_0^4)/(4-3h_0+0.5h_0^3); h_0 \le 1 \nonumber \\
	&&\langle sin\gamma \rangle =\langle sin\gamma (h_0=1) \rangle =0.53;h_0>1
\end{eqnarray}
At lower $h_0$-values (lower operation temperatures) smaller angles dominate (see Fig. \ref{fig:fig9}). For values $h_0>1$ the mean scattering angle stays constant, as only $h$-values below 1.0 allow for electron losses. 
Therefore, and by finding an optimum match with measurements the following combination of F–functions for specular and diffuse reflection is proposed: 
\begin{subequations}
	\begin{equation}
		F(h_0)=(a-\eta b)\cdot F_{s1}+(1-a+\eta b)F_{d1}, \text{for} \quad h_0<0.5 \label{eq-F_function_1}
	\end{equation}
	\begin{equation}
		\text{with} \quad 2a-b=1, \eta=\frac{\langle sin\gamma(h_0) \rangle-\langle sin\gamma  (h_0=0) \rangle}{\langle sin\gamma(h_0=1.0) \rangle -\langle sin\gamma (h_0=0) \rangle}
	\end{equation}
	\begin{equation}
		F(h_0)=(a-\eta b)\cdot F_{s2}+(1-a+\eta b)F_{d1}, \text{for} \quad 0.5\le h_0<1  \label{eq-F_function_2}
	\end{equation}
	\begin{equation}
		F(h_0)=(a-b)\cdot F_{s2}+(1-a+b)F_{d2}, \text{for} \quad h_0>1 \label{eq-F_function_3}
	\end{equation}
\begin{equation}
	\text{An adequate choice is} \quad a=0.75, \quad b=0.5
\end{equation}
\end{subequations}

The F-functions from Eqs. (\ref{eq-F_function_1}, \ref{eq-F_function_2} and \ref{eq-F_function_3}) are to be used finally in the recursion Eq. \ref{eq-recursion_process}.

\begin{figure}[!h]
	\centering
	\includegraphics[width=\columnwidth]{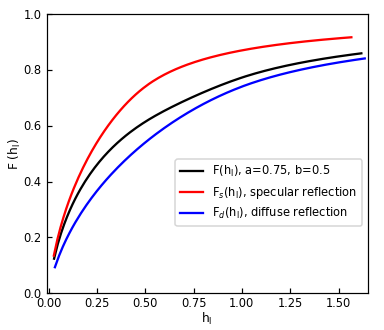}
	\caption{\label{fig:fig10}Plot of $F_s (h_l )$, $F_d (h_l )$, $F(h_l )$, for the mixing parameters a=0.75, b=0.5.}
\end{figure}

Fig. \ref{fig:fig10} allows a comparison of the three functions. An easy transformation from DC- to RF-conductivity and vice-versa is possible by Table \ref{tab:tableA}. The tabulated values refer to the matched F - function with $a=0.75$. 

Alternatively, fitting functions Eqs. (\ref{eq:eq_A5}), (\ref{eq:eq_A6}) for $F_l(h_0)$ and $F_l(q_0)$ for copper are offered for the experimentally found mixing parameter  $a=0.75$. This allows to directly get in good approximation the RF conductivity $\sigma_a$ from DC - values, or vice-versa the effective DC conductivity from measured Q - values at given temperatures.

As an application example, the frequency dependence of $\sigma_a(T)$ is calculated in the following way:
\begin{itemize}
	\item[-] calculate $h_0$ for a given DC conductivity $\sigma(T)$ and operation frequency $f$ by Eq. (\ref{eq-define_h0}),
	\item[-] get the related $F_l$-function directly from Table \ref{tab:tableA} (or from the fitting function (\ref{eq:eq_A5})).
\end{itemize}

\begin{figure}[!h]
\includegraphics[width=\columnwidth]{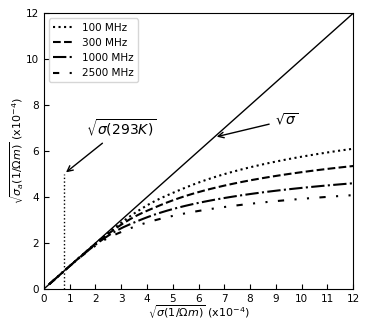}
\caption{\label{fig:fig18} Calculation of $\sqrt{\sigma_a}$ over $\sqrt{\sigma}$ for different frequencies with the geometric model for copper.}
\end{figure}

Figure \ref{fig:fig18} shows the result by plotting $\sqrt {\sigma_a}$ against $\sqrt {\sigma}$ for four different frequencies and using the interpolated function $F_l$ from the last column of Table \ref{tab:tableA}.

\subsection{Specific Surface Resistance}
The wall losses in the cavity at a surface element $dxdz$ with magnetic surface field amplitude $H_0$ are given by
\begin{equation}
	dP=\frac{1}{2}H_0^2\cdot R\cdot dxdz \label{eq:eq38}
\end{equation}
with the specific surface resistance, by including anomalous losses:   
\begin{eqnarray}
	R(T)/\Omega&&=\frac{\rho_a(T)}{\delta_a(T)}=\sqrt{\frac{\omega\cdot \mu_0}{2\sigma_a(T)}}=\sqrt{\frac{\omega\cdot \mu_0}{2\sigma(T)\cdot F_l(h_0)}} \nonumber \\
	&&=R_c(T)\cdot F_l(h_0)^{-1/2}. \label{eq:eq39}
\end{eqnarray}
$R_c(T)$ is the “classic” result regarding the normal skin effect only.

This means, the surface resistance by including the anomalous skin effect is a factor $F_l(h_0)^{-1/2}$ larger than predicted from the normal skin effect. Eq. (\ref{eq:eq39}) is applicable, as soon as the DC–conductivity values $\sigma(T)$ are known to calculate $h_0(\sigma)$ and to insert the belonging, tabulated $F_l(h_0)$–value.

For comparison, the same surface ratio can be calculated from the formulas as published by R.G. Chambers\cite{chambers1952anomalous} with the ASE diffusion model. The interpolation formula in \cite{chambers1952anomalous}, Eq. (A.1) and inserting the constants in $R_{\infty}$ gives for copper
\begin{subequations}
	\begin{equation}
		R(T)=1.123\cdot 10^{-9}\cdot f^{2/3}[1+F_R\cdot (\alpha)^{-G}].  \label{eq:eq40a}
	\end{equation}
\text{Under the assumption of diffuse electron reflection at}
\text{the surface the coefficients are}
\begin{equation}
	F_R=1.157, \quad G=0.2757.  \label{eq:eq40b}
\end{equation}
\text{In case of specular reflection instead, the formula from}
\text{R.G. Chambers results in the coefficients}
\begin{equation}
	F_R=1.376, \quad G=0.3592.  \label{eq:eq40c}
\end{equation}
\text{When applying the same formula to the results of Reu-}
\text{ter-Sonderheimer\cite{reuter1948theory}, Chambers found for that case the}
\text{adequate coefficients}                                        
\begin{equation}
	F_R=1.004, \quad G=0.333.  \label{eq:eq40d}
\end{equation}
\end{subequations}
The “classic” case, neglecting the anomalous skin effect gives 
\begin{equation}
    R_c(T)=\sqrt{\pi f \mu_0 \rho (T)}. \label{eq:eqRc}
\end{equation}

The parameter $\alpha$ in Eq. (\ref{eq:eq40a}) is related to $h_0$ as defined above by $\alpha=1.5\cdot (h_0)^{-2}$ - see Eq. (\ref{eq-define_h0}) and Ref. \cite{reuter1948theory} - Eq. (19). This gives then with Eq. (\ref{eq:eq40a} and \ref{eq:eqRc})
\begin{equation}
	\frac{R(T)}{R_c(T)}=\frac{[1+F_R\cdot (1.5/h_0^2)^{-G}]}{1.93\cdot h_0^{1/3}}.  \label{eq:eq40e}
\end{equation}

\begin{figure}
     \begin{subfigure}[!h]{\columnwidth}
         \includegraphics[width=8cm]{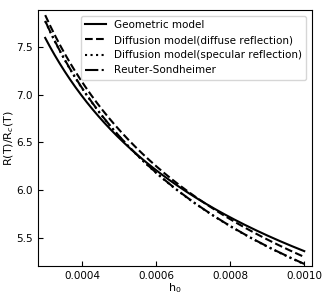}
         \label{fig:fig11a}
     \end{subfigure}
     \begin{subfigure}[!h]{\columnwidth}
         \includegraphics[width=8cm]{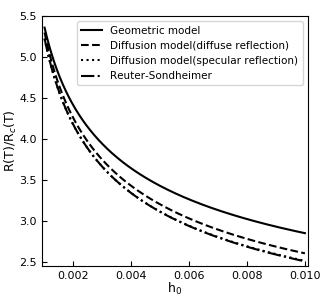}
         \label{fig:fig11b}
     \end{subfigure}
     \begin{subfigure}[!h]{\columnwidth}
         \includegraphics[width=8cm]{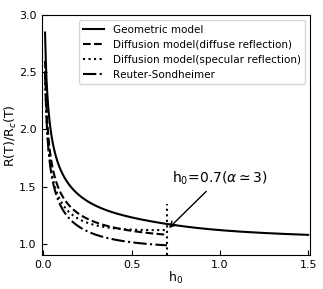}
         \label{fig:fig11c}
     \end{subfigure}
        \caption{\label{fig:fig11}Plot of $R(T)/R_c(T)$ according to Eq. (\ref{eq:eq39}) and to Eqs. (\ref{eq:eq40a}, \ref{eq:eq40b}, \ref{eq:eq40c}, \ref{eq:eq40d}, \ref{eq:eq40e}) for $0.0003\leq h_0 \leq 0.001$, $0.001\leq h_0 \leq 0.1$, $0.01\leq h_0 \leq 1.5$, respectively.}
\end{figure}

According to Ref. \cite{chambers1952anomalous}, the equations of the diffusion model are applicable for $\alpha>3$, that is $h_0<0.7$, where the anomalous skin effect makes significant contributions. As can be seen by Fig. \ref{fig:fig11}, the geometric model predicts a larger contribution of the anomalous skin effect than all diffusion model variants as soon as $h_0>0.0005$. The main differences between both models occur, when the skin depth and the free path length are of the same order of magnitude.

\begin{figure}[!h]
\includegraphics[width=\columnwidth]{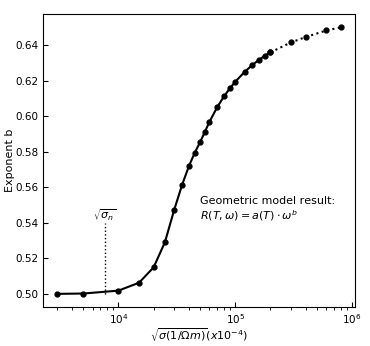}
\caption{\label{fig:R_omega} The exponent $b$ of $\omega$ as resulting from the geometric model for copper.}
\end{figure}

Another comparison between models is the dependence of $R(T)$ on $\omega$ at a given temperature. The diffusion model predicts a transition from $\omega^{1/2}$ towards $\omega^{2/3}$ in the strongly ASE dominated region. Eq. (\ref{eq:eq39}) is rewritten in the form
\begin{equation}
    R(T, \omega)=a(T)\cdot \omega^{b} \label{eq:R_omega}
\end{equation}
Figure \ref{fig:R_omega} shows the result for the dependence of $b$ from $\sqrt{\sigma(T)}$: \\
It continuously changes from 0.5 at low conductivity values towards 0.64 and above in the extreme ASE regime - in accordance with the diffusion ASE model and with experiments.

\section{CAVITY PREPARATION AND RF MEASUREMENT RESULTS }
\subsection{Overview}
A main intention was the experimental investigation of the anomalous skin effect at RF frequencies relevant for ion acceleration. From thermal investigations with respect to cavity cooling at cryogenic temperatures it was clear, that bulk copper cavities are a first choice. This was the reason to build a coaxial test cavity from bulk copper. After first measurements it was noted, that mass ware SF copper as used in electrotechnics and many other fields can have rather low RRR - values. In a next step, the surface of the 340 MHz cavity was electrolytically matt copper-plated by company Galvano-T with a layer thickness of 100 $\mu m$. After encouraging measurements, annealing for one hour at 400°C as suggested in Ref. \cite{fouaidy2006rrr} resulted in the final cavity performance. 

\subsection{Cavity layout and construction}
\begin{figure}[!h]
\includegraphics[width=8cm]{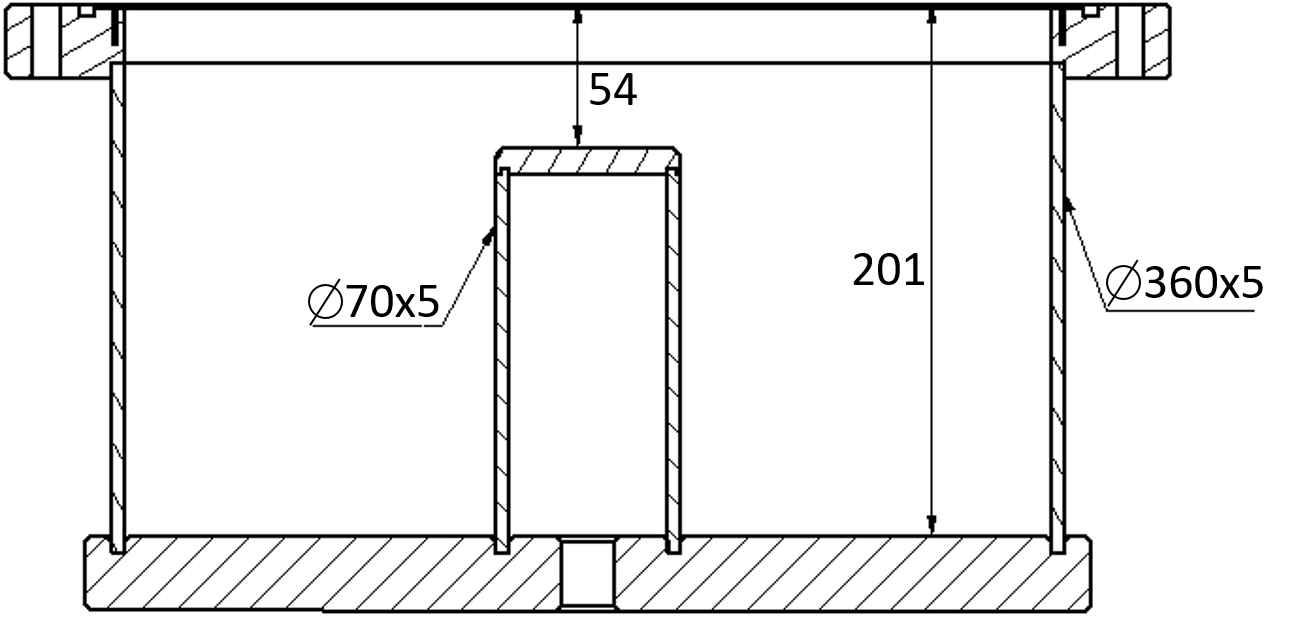}
\caption{\label{fig:fig12} Cross-sectional view of the 340 MHz cavity with main dimensions.}
\end{figure}

End flanges from bulk copper were produced and the tubes were in a one-step procedure vacuum-brazed with the end-plates by a silver-solder (Fig. \ref{fig:fig12}).

\begin{figure}
     \begin{subfigure}[!h]{\columnwidth}
         \includegraphics[width=7cm]{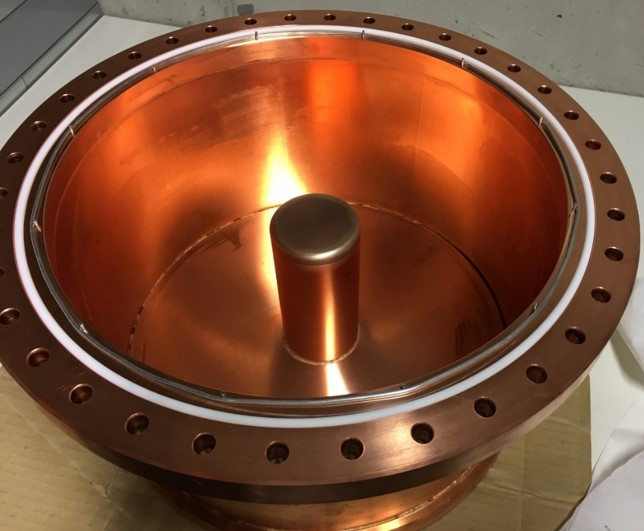}
         \label{fig:fig13a}
     \end{subfigure}
     \begin{subfigure}[!h]{\columnwidth}
         \includegraphics[width=7cm]{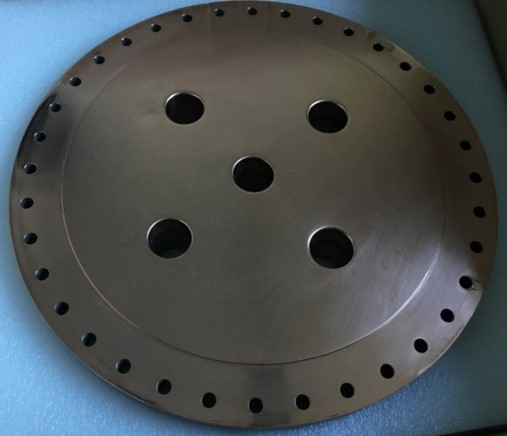}
         \label{fig:fig13b}
     \end{subfigure}
        \caption{Photo of the 340 MHz cavity with lid removed (top) and of the lid inner surface after copper plating (bottom). The RF contact to the lid is provided by a 2 mm aluminum wire, backed by a Teflon O-ring.}
        \label{fig:fig13}
\end{figure}

The end-lid at the gap-side was manufactured from stainless steel and was galvanically copper-plated. This allowed to apply the well-proven CF-sealing flange technology: all flanges of the cavity were placed on that end-lid. This end-lid was kept in the original state for all measurements. CST-simulations show, that the power loss on the end-lid contributes with 5.3\%. As this component was not included in the annealing process, the final Q-values on the 340 MHz cavity may be up to 2\% lowered by that effect at the low temperatures. Figure \ref{fig:fig12} indicates the main dimensions of the 340 MHz cavity, and photos of the same cavity are shown by Fig. \ref{fig:fig13}.  

\subsection{RF Measurements}
Before every cold-test the cavity was RF-conditioned up to 50 W in CW-mode to clean the surfaces and to transit the lowest multipacting levels. A first measurement at room temperature gave the results as shown by Table \ref{tab:table2}. At the same time, the theoretical numbers from CST–simulations are given for comparison ($\rho$ was assumed to be $1.7\cdot 10^{-8}\Omega m$).
\begin{table}[!h]
\caption{\label{tab:table2}
Characteristic cavity parameters at room temperature.}
\begin{ruledtabular}
\begin{tabular}{cccc}
$f_{cavity}$ & $Q_{0,sim}$ & $Q_{0,meas}$ & $Q_{0,meas}/Q_{0,sim}$ \\
\hline
340 MHz&	19665&	17569&	0.893\\
\end{tabular}
\end{ruledtabular}
\end{table}

\begin{figure}[!h]
\includegraphics[width=\columnwidth]{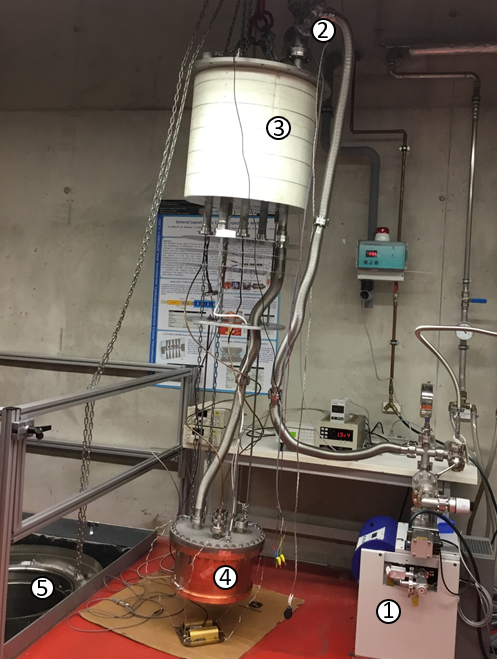}
\caption{\label{fig:fig14} Photo of the prepared 340 MHz cavity, ready for installation in the cryostat: 1=turbo molecular pump station, 2=ion getter pump, 3=lid of the vertical cryostat, 4=cavity, 5=open cryostat.}
\end{figure}

The cavity was tested in a vertical He-cryostat. The prepared 340 MHz cavity with cryostat-lid at the top is shown by Fig. \ref{fig:fig14}. The cavity was pumped by a combination of a turbo pumping-station and an ion getter-pump, as used routinely in this laboratory for the test of superconducting cavities.

\begin{figure}[!h]
\includegraphics[width=\columnwidth]{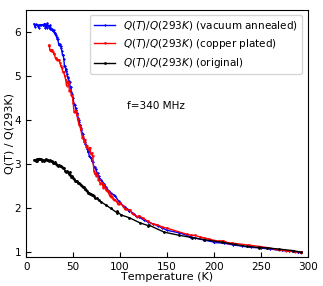}
\caption{\label{fig:fig15} Measured $Q$ ratios over T for all three surface conditions of the 340 MHz cavity.}
\end{figure}

The resulting $Q(T)/Q_n$ - graph for the 340 MHz cavity is the black curve in Fig. \ref{fig:fig15}. From this result and later studies it became evident, that the original RRR - values of the used copper-components were very low - RRR $\cong$ 19. This is not unusual for bulk-copper with unspecified cryogenic capabilities.

In a next step the 340 MHz bulk copper cavity (without the stainless-steel end-lid) was matt copper-plated by use of the commercial plating chemical Cuprostar LP1, which gives a crystalline, silky matt deposition\cite{szcepaniak}. This method was chosen, as it applies less organic ingredients in the copper–bath when compared to high-gloss plating. The measurement for that case (red curve in Fig. \ref{fig:fig15}) was not as stable as the other two: A weak cable connection within the cryostat was found as the reason afterwards. Finally, inspired by Ref. \cite{fouaidy2006rrr}, vacuum-annealing at 400°C for one hour of the copper-plated copper cavity was performed in the vacuum furnace. The resulting values are plotted in Fig. \ref{fig:fig15} (blue curve) and indicate, that this treatment indeed is giving some improvement additionally to the copper-plating.

\subsection{Comparison of measurements with theory}
From the measurements of $Q(T)$, depending on $\sigma_a(T)^{1/2}$, one can now deduce the classical result $Q_c(T)$ from the normal skin effect and the effective DC conductivity $\sigma(T)$ with the geometric model in the following way:
\begin{itemize}
	\item[-] $q_0$ is obtained by inserting $\frac{\sigma_n}{\sigma_a(T)}=\left( \frac{Q(293K)}{Q(T)} \right) ^2$ into Eq. \ref{eq-define_q0}
	\item[-] the related $q_l$ and $F(q_l)$ is given by Table \ref{tab:tableA} (or from the recursion process as described in Eqs. (\ref{eq-reverse_recursion})) as a result of the geometric model
    \item[-] $\sigma(T)=\sigma_a(T)/ F(q_l)$
\end{itemize}

\begin{figure}[!h]
\includegraphics[width=\columnwidth]{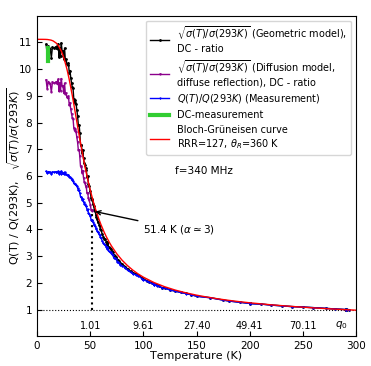}
\caption{\label{fig:fig16} Square-roots of the DC conductivity ratios as calculated from the measured Q-value ratios. The DC measurement for galvanically copper-plated layers (in green) is from Ref. \cite{fouaidy2006rrr}.}
\end{figure}

Moreover, a comparison with Eqs. (\ref{eq:eq40a}, \ref{eq:eq40b}) from Ref. \cite{chambers1952anomalous} is made. The results for the vacuum annealed cavity case are plotted in Fig. \ref{fig:fig16}. The resulting ratio of $\sqrt{\sigma(T)/\sigma(293K)}\cong 11$ at lowest temperatures $T \le 25K$ means, that the RRR - value for that copper sheath corresponds to RRR $=11^2\cong 121$. This is just slightly above the results published in Ref. \cite{fouaidy2006rrr} for DC measured RRR - values between 107 and 117 of galvanically copper-plated and annealed layers. A DESY - investigation on RRR - values on the same electrolytic bath from Galvano-T as used for our measurements results in RRR values between 110 and 117\cite{ermakov2014}. The corresponding value from the diffusion model is about 15\% below those DC measurements. This shows again, that the anomalous skin effect is underestimated by the diffusion model at these frequencies and temperatures.

\begin{figure}[!h]
\includegraphics[width=\columnwidth]{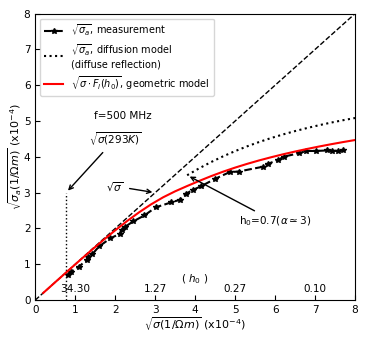}
\caption{\label{fig:fig17} $Q$–measurements on a 500 MHz cavity\cite{weingarten} and resultant $\sqrt{\sigma_a}$-values against $\sqrt{\sigma}$. Comparison with the geometric model prediction gives good agreement, while the diffusion model (Eqs. \ref{eq:eq40b}, \ref{eq:eq40e}) predicts $Q$-values, which are about 16\% higher.}
\end{figure}

Some published measurements are also compared with the prediction from the geometric model as presented in this paper. Ref. \cite{weingarten} shows a measurement of $\sqrt{\sigma_a}$ (deduced from $Q$-measurement) over $\sqrt{\sigma}$ on a 500 MHz cavity. The geometric model fits these data very well, as shown by Fig. \ref{fig:fig17}. 

\begin{figure}[!h]
\includegraphics[width=\columnwidth]{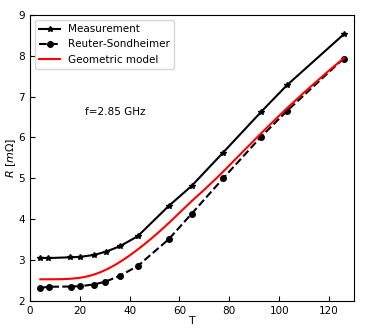}
\caption{\label{fig:fig19} Surface resistance $R(T)$ for 2.85 GHz cavity measurements at SLAC, Stanford \cite{CahillIPAC2016-MOPMW038} and comparison with the predictions from the geometric model when fed with DC conductivities from a Bloch-Grüneisen curve with RRR = 400 and $\theta_R$=340 K\cite{matula1979electrical}.}
\end{figure}

Finally, the geometric model was also compared with measurements at a 2.85 GHz cavity, as published in Ref. \cite{CahillIPAC2016-MOPMW038} - Fig. 3.22. This cavity was made from copper with RRR=400. In that case, the geometric model predictions are again closer to the measurements than the diffusion model, as shown by Fig. \ref{fig:fig19}. It is in general assumed that the needed efforts in surface preparations for reaching theoretical predictions become harder with increasing RF frequency.

\section{Cavities for PULSED ION LINACS AT CRYOGENIC TEMPERATURES}
From the measurements described in the previous chapter, one can conclude that the RF conducting surface has to be electrolytically copper-plated in case that the RRR  – value of the used copper was not specified. This concept can deliver RRR - values around 120. Industrial offers for semi-finished products from OFHC–copper up to RRR = 400 are on the market. On the other hand, Fig. \ref{fig:fig1} clearly shows, that RRR - values above 200 will not improve the results significantly, as long as the intended operation temperature is 40 K or higher.

The bulk cavity material has to provide excellent heat conduction properties to keep the cavity surface close to the temperature of the coolant. Copper-plated steel cavities are definitely excluded for this application. 

The electrical conductivity dependence on temperature limits the allowed surface temperature increase within an RF pulse:\\
From Eqs. (\ref{eq:eq38}, \ref{eq:eq39}) it follows that $P\sim \sigma_a^{-1/2}$. If for example a change of the amplifier RF power of 5\% during one pulse is assumed acceptable and can be handled safely by the low-level RF controls, one can find from the function $\sqrt{\sigma_a(T)}$ the allowed temperature increase $\Delta T$, corresponding to a 5\% - variation of $\sqrt{\sigma_a(T)}$.

In the following, RF voltage levels, corresponding to a power loss density of $5 MW/m^2$ at 300 K are investigated at 40 K. The surface temperature rise and the temperature profile into the bulk copper is studied for RF pulse lengths $\tau$ of 50, 200 and 500 $\mu s$. The power level at 40 K operation is reduced according to the corresponding change of $Q_0$, to keep the acceleration voltage for both cases about constant.

\subsection{Surface temperature response and temperature profile}
Assuming an instantaneous pulse of energy $W_0=p\cdot \tau \cdot A$ on the cavity surface $A$ in the $xz$-plane, the temperature response in time and along $y$ into the copper wall after the pulse, is given by Ref. \cite{holman2010heat} as
\begin{equation}
    \Delta T = \frac{W_0}{A\rho c \cdot \sqrt{\pi \alpha t}}\cdot e^{-y^2/4\alpha t} 
\end{equation}
with $W_0/J$, $c/kJ/kg\cdot K$, $\alpha/m^2/s$, $\rho/kg/m^3$, $A/m^2$. $t$ = time after heat pulse. $y=0$ marks the surface
plane.

\begin{figure}[!h]
     \begin{subfigure}[htbp]{\columnwidth}
         \includegraphics[width=7.5cm]{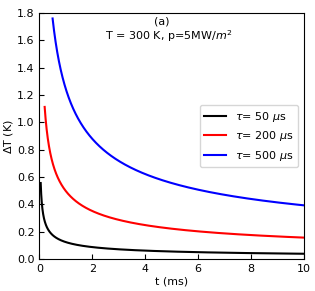}
     \end{subfigure}  
     \begin{subfigure}[tbph]{\columnwidth}
         \includegraphics[width=7.5cm]{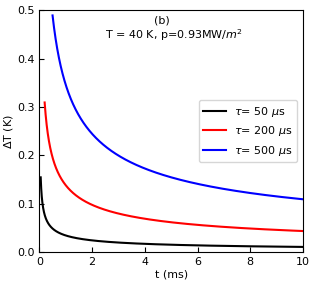}
     \end{subfigure} 
        \caption{Plot of $\Delta T(t)$ for three pulse lengths and at two operation temperatures.}
        \label{fig:fig20}
\end{figure}

For the values of the specific heat $c$ as well as for the thermal diffusivity $\alpha$ a wall from bulk copper with RRR = 100 at the according starting temperatures was assumed as a realistic case. The results for the time dependence of the surface temperature at $y=0$ are plotted in Fig. \ref{fig:fig20} for two different operating temperatures. 

One can see, that pulse lengths as needed for injection into proton and ion synchrotrons can be realized by a cryogenic operation of copper linacs, up to very high acceleration fields. The temperature rise during one pulse remains within an acceptable range at a defined 40K - operation temperature:\\
From measurements as shown by Fig. \ref{fig:fig15} it can be deduced that the relative change of the intrinsic $Q$-value is about 1.7\%/K at that temperature. This will lead to only a modest change in reflected RF power during pulses like characterized in Fig. \ref{fig:fig20} (b). Pulse lengths up to the $ms$ - range seem feasible.

\begin{figure}[!h]
     \begin{subfigure}[htbp]{\columnwidth}
         \includegraphics[width=7.5cm]{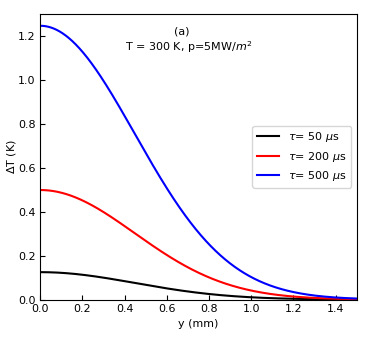}
     \end{subfigure}
     \begin{subfigure}[htbp]{\columnwidth}
         \includegraphics[width=7.5cm]{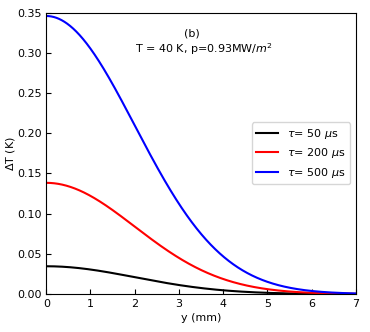}
     \end{subfigure}
        \caption{Plot of $\Delta T(y)$ one $ms$ after the energy was applied on the surface, for two starting temperatures.}
        \label{fig:fig21}
\end{figure}

\begin{figure}[!h]
     \begin{subfigure}[htbp]{0.85\columnwidth}
         \includegraphics[width=\columnwidth]{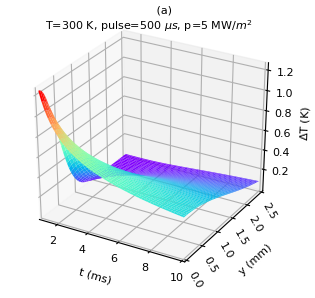}
     \end{subfigure}
     \begin{subfigure}[htbp]{0.85\columnwidth}
         \includegraphics[width=\columnwidth]{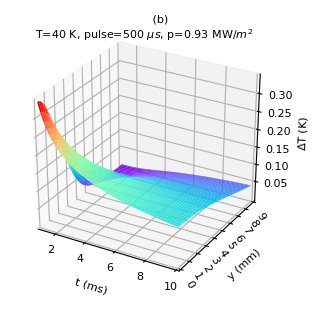}
     \end{subfigure}
        \caption{Temperature distribution in time and penetration depth for two starting temperatures.}
        \label{fig:fig22}
\end{figure}

Another important aspect is the heat transport velocity from the cavity surface to the coolant. It has to be assured, that the cavity surface temperature has come back close to the starting value before the next RF pulse would be applied. Figure \ref{fig:fig21} shows the temperature profiles along the heat transport coordinate $y$ – one $ms$ after the heat pulse application, $t=1 ms$. Figure \ref{fig:fig22} finally shows the two-dimensional temperature distributions along $t$ and $y$ for both operation-temperatures.

\subsection{Suggested cavity geometries and cooling}
\begin{figure}[!h]
    \includegraphics[width=\columnwidth]{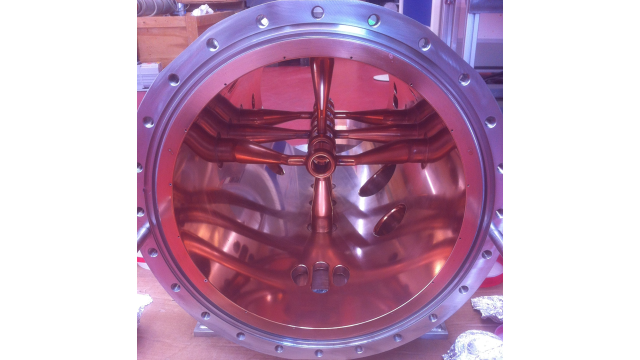}
    \caption{View into a 7 gap, 325 MHz prototype CH–cavity for room-temperature operation\cite{AlmomaniIPAC2014}. The inner diameter is 330 mm.}
    \label{fig:fig23}
\end{figure}

Compact and simple drift tube cavities like shown by Fig. \ref{fig:fig23} are preferable for cryogenic temperature operation. KONUS- and APF-beam dynamics allow for a sequence of simple drift tubes, containing no focusing elements. In case of KONUS, magnetic quadrupole triplets are providing the transverse focusing after each sequence of simple drift tubes\cite{ratzinger2019combined}. The cavities should preferably be built from bulk copper and with RRR - values not lower than about 150 to get an attractive performance at operation temperatures between 40 and 50 K. H-type cavities seem very well suited for this purpose. However, the transition from the established copper-plated stainless-steel cavity technology to bulk copper cavities has to be made. An alternative might be thick-plated (a few mm) stainless steel cavities with the coolant flowing in the copper sheath.

Cooling down to 77 K is easy with liquid nitrogen. The final cooling stage has to be designed according to the exact operating temperature and size of a facility. Optional concepts are described in Ref. \cite{dalakov2020innovative}.

\section{SUMMARY AND OUTLOOK}
The geometric model was presented, which is adequate to describe the extra power losses by the anomalous skin effect at RF frequencies up to about 2.5 GHz. This is the range most commonly used for ion acceleration. Measurements on a 340 MHz cavity from bulk copper with a very low original RRR-value showed promising RF results between 40 K and 50 K after electrolytical copper-plating and vacuum-annealing at 400°C. Applying the geometric ASE model leads to realistic results for surface-conductivities, deduced from known DC conductivities and vice versa. Experimental data at 340 MHz, 500 MHz and at 2.85 GHz from literature are described very well by the geometric model. Tabulated data as well as two fitting functions are presented in the appendix for direct application of the model on copper cavities.

In a next step, measurements on the temperature behavior of cavities made from specified copper are under preparation, combined with voltage breakdown tests at 433 MHz. 

It looks promising to develop pulsed linacs like ion synchrotron injectors and medical linacs in cold copper technology. When being operated at temperatures around 40 K they show great potential for a realization by using compact and efficient cavities of the H-type.

\begin{acknowledgments}
The GSI support by giving access to their mechanical infrastructure is highly appreciated as well as the high motivation of the IAP mechanical workshop and of the company Galvano-T, Windeck, Germany. Especially, we like to thank Hendrik Hähnel, Holger Podlech and Giuliano Franchetti for support and fruitful discussions.
\end{acknowledgments} 


\appendix

\section{Conducting electrons in velocity space} 
\subsection{Fraction of lost RF conduction electrons in a $y$-plane within the skin sheath}\label{app:density_y_plane}
The current density with the loss due to the anomalous skin effect in the velcocity range $h \le w \le 1, h<1$:

\begin{widetext}

\begin{subequations}
	\begin{eqnarray}
		j_{z,1}^{\prime}(h)\sim e\cdot 2\Delta v_F\cdot v_F^3\cdot \int_{h}^{1}\int_{-(1-w^2)^{1/2}}^{(1-w^2)^{1/2}}(1-u^2-w^2)^{1/2}\cdot \frac{h}{w+h} dudw,
	\end{eqnarray}
	\begin{eqnarray}
		j_{z,1}^{\prime}(h) \sim e\cdot \Delta v_F\cdot v_F^3 \cdot \pi \left[ -\frac{h^3}{2} +h^2-\frac{h}{2}+(h-h^3) \cdot ln(h+1)-(h-h^3)\cdot ln(2h) \right]. \label{eq-j_z1}
	\end{eqnarray}
	\text{The loss-free part within $ 0 \le w < h, h<1$ is}
	\begin{eqnarray}
		j_{z,2}^{\prime}(h)\sim e\cdot 2\Delta v_F\cdot v_F^3\cdot \int_{0}^{h}\int_{-(1-w^2)^{1/2}}^{(1-w^2)^{1/2}} (1-u^2-w^2)^{1/2} dudw,
	\end{eqnarray}
	\begin{equation}
		j_{z,2}^{\prime}(h) \sim e\cdot \Delta v_F\cdot v_F^3 \cdot \pi \left[ h-\frac{h^3}{3} \right] \label{eq-j_z2}
	\end{equation}
	\text{The completely loss-free zones in case of $h\ge 1$ are contributing to the current density as}
	\begin{eqnarray}
		&&j_{z,3}^{\prime}(h)\sim e\cdot 2\Delta v_F\cdot v_F^3\cdot \int_{0}^{1}\int_{-(1-w^2)^{1/2}}^{(1-w^2)^{1/2}} \nonumber \\
  &&(1-u^2-w^2)^{1/2} dudw=\frac{2}{3}\pi e \Delta v_F v_F^3 \label{eq-j_z3}
	\end{eqnarray}
\end{subequations}
(Note: The factor 2 difference against Eq. (\ref{eq:eq8}) results from the integration on positive $w$ only.)

\subsection{Reduced RF conductivity within the skin sheath}\label{app:density_whole_sheat}

\textbf{Assumption 1, Specular reflection, $p=1$}:\\
Two cases have to be distinguished: \\
Case 1: For $2h_0<1$ losses will occur for all $h$ and the integration is performed over the whole $h$-range: \\
\begin{subequations}
	\begin{eqnarray}
		j_{z,a1}\sim e\cdot \pi \cdot \Delta v_F\cdot v_F^3\cdot \frac{1}{h_0} \cdot \left[ \int_0^{2h_0}j_{z,1}^\prime(h)\cdot dh+\int_0^{2h_0}j_{z,2}^\prime(h)\cdot dh \right] 
	\end{eqnarray}
	\text{The result is}
	\begin{equation}
		j_{z,a1}\sim  \frac{4\pi}{3}\cdot e \cdot v_F^3 \cdot \Delta v_F \cdot F_{s1}(h_0)
	\end{equation}
\text{with}
	\begin{equation}
		F_{s1}(h_0)=0.375-0.95445h_0+1.5h_0^2+1.6587h_0^3 +ln(1+2h_0)\cdot(1.5h_0-3h_0^3-0.1875h_0^{-1})+ln(h_0)\cdot(3h_0^3-1.5h_0) \label{eq:eq23b}
	\end{equation}
\text{Case 2: $2h_0 \ge 1$}
	\begin{equation}
		j_{z,a1}\sim e\cdot \pi \cdot \Delta v_F\cdot v_F^3\cdot \frac{1}{h_0}\left[ \int_0^{1}j_{z,1}^\prime(h)\cdot dh +\int_0^{1}j_{z,2}^\prime(h)\cdot dh+\int_{1}^{2h_0}j_{z,3}^{\prime}(h)\cdot dh \right],
	\end{equation}
\text{This results in}
	\begin{equation}
		j_{z,a1}\sim  \frac{4\pi}{3}\cdot e \cdot v_F^3 \cdot \Delta v_F \cdot F_{s2}(h_0)
	\end{equation}
\text{with}
	\begin{equation}
		F_{s2}(h_0)=1.0-0.1299\cdot h_0^{-1} \label{eq-fs2}
	\end{equation}
\end{subequations}
\textbf{Assumption 2, Diffuse reflection, $p=0$}: \\
Case 1, $h_0<1$: \\
\begin{subequations}
	\begin{equation}
		j_{z,a1}\sim e\cdot \pi \cdot \Delta v_F\cdot v_F^3\cdot \frac{1}{h_0}  \cdot \left[ \int_0^{h_0}j_{z,1}^\prime(h)\cdot dh+\int_0^{h_0}j_{z,2}^\prime(h)\cdot dh \right] ,
	\end{equation}
	\begin{equation}
		j_{z,a1}\sim  \frac{4\pi}{3}\cdot e \cdot v_F^3 \cdot \Delta v_F \cdot F_{d1}(h_0)
	\end{equation}
	\text{with}
	\begin{equation}
		F_{d1}(h_0)=0.375+0.0426h_0+0.375h_0^2-0.0525h_0^3 -ln(1+h_0)\cdot(0.375h_0^3-0.75h_0+0.375h_0^{-1}) +ln(h_0)\cdot (0.375h_0^3-0.75h_0) \label{eq-fd1}
	\end{equation}
	\text{Case 2: $h_0 \ge 1$}
	\begin{equation}
		j_{z,a1}\sim e\cdot \pi \cdot \Delta v_F\cdot v_F^3\cdot \frac{1}{h_0}
		\left[ \int_0^{1}j_{z,1}^\prime(h)\cdot dh  +\int_0^{1}j_{z,2}^\prime(h)\cdot dh+\int_{1}^{h_0}j_{z,3}^{\prime}(h)\cdot dh \right],
	\end{equation}
	\text{This results in}
	\begin{equation}
		j_{z,a1}\sim  \frac{4\pi}{3}\cdot e \cdot v_F^3 \cdot \Delta v_F \cdot F_{d2}(h_0)
	\end{equation}
	\text{with}
	\begin{equation}
		F_{d2}(h_0)=1.0-0.2598\cdot h_0^{-1} \label{eq-fd2}
	\end{equation}
\end{subequations}

\section{Table and Fitting Curves to the Geometric Model for Copper} \label{sec-appendix_b}
Chapter \ref{chap:chap2E} derives the relation of the RF conductivity $\sigma_a(T)$ against the DC conductivity $\sigma(T)$ by defining the functions $F_l (h_0)$ and $F_l (q_0)$. The recursion process to derive these functions was described in that chapter. In the following, these functions are given in tabulated form and approximately as fitting equations. The experimentally justified combination of specular and diffuse reflection as described by Eqs. (\ref{eq-F_function_1}, \ref{eq-F_function_2} and \ref{eq-F_function_3}) is used here. A description how to apply the fitting curves or Table \ref{tab:tableA} alternatively, is given in the following.

\vspace{\baselineskip}
Case 1: The anomalous conductivity $\sigma_a(T)$ is to be derived from a known DC conductivity $\sigma(T)$: 
\begin{itemize}
	\item[$\bullet$] Calculation of \quad \quad \quad \quad \quad $h_0=\frac{7.7\cdot 10^{17}}{(\sigma(T))^{3/2}\cdot f^{1/2}}$ 
	\item[$\bullet$] Calculation of $\sigma_a(T)$ by \quad $\sigma_a(T)=F_l(h_0)\cdot \sigma(T)$ 	
\end{itemize}

$F_l (h_0)$ can be calculated approximately by the fitting curve defined by Eq. (\ref{eq:eq_A5}), or taken from Table \ref{tab:tableA}. The specific surface resistance is then given by inserting $F_l (h_0)$ into Eq. (\ref{eq:eq39}).
\begin{equation}
    F_l(h_0)=a_1(a_2\cdot h_0)^{a_3}+b_1\cdot b_2^{h_0+b_3} +c_1\cdot ln(c_2\cdot h_0+c_3)+d  \label{eq:eq_A5}
\end{equation}
With \\
$a_1=1.266851$, $a_2=0.984127$, $a_3=0.454241$,
$b_1=0.953263$, $b_2=0.765229$, $b_3=-1.092828$,
$c_1=59.722208$, $c_2=-0.001984$, $c_3=0.862839$, 
$d=7.508484$.

\vspace{\baselineskip}
Case 2: If $Q(T)$ – values have been measured, one gets the according $\sigma_a(T)$ – values from Eq. (\ref{eq:eqQVgA}), which says that $\sigma_a(T)\propto (Q(T))^2$. When $Q_n$ and $Q(T)$ were measured, one gets $\sigma_a(T)$ by $\sigma_a(T)= \sigma_n\cdot (Q(T)/Q_n)^2$. The DC conductivity $\sigma(T)$ is then derived by the following way:
\begin{itemize}
	\item[$\bullet$] Calculate $q_0(T)$ by Eq. (\ref{eq-define_q0}): $q_0=\frac{7.7\cdot 10^{17}}{(\sigma_a(T))^{3/2}\cdot f^{1/2}}$ 
	\item[$\bullet$] Calculation of $\sigma(T)$ by \quad \quad $\sigma(T)=\sigma_a(T)/F_l(q_0)$ 	
\end{itemize}

$F_l (q_0)$ can be calculated approximately by the fitting curve defined by Eq. (\ref{eq:eq_A6}), or taken from Table \ref{tab:tableA}.
\begin{equation}
    F_l(q_0)=a_1(a_2\cdot q_0)^{a_3}+b_1\cdot b_2^{q_0+b_3}+c_1\cdot ln(c_2\cdot q_0+c_3)+d_1\cdot cos(d_2\cdot q_0+d_3)+m  \label{eq:eq_A6}
\end{equation}
With\\
$a_1=12.460768$, $a_2=10.486607$, $a_3=0.991990$, 
$b_1=117.007522$, $b_2=0.946872$, $b_3=-44.355124$, 
$c_1=260.470406$, $c_2=-2.247174$, $c_3=0.000819$, 
$d_1=-1768.908809$, $d_2=0.046725$, $d_3=3.765885$, 
$m=-901.963777$.

The fitting curves and comparison with calculated points from the recursion formulas are shown by Fig. \ref{fig:fitting_plots}.

\begin{table}[htbp]
\centering
\caption{\label{tab:tableA}Values of $F_l$ according to $h_0$, $(h_l=q_l)$, $q_0$, valid for copper.}
\begin{subtable}{0.45\linewidth}
\begin{tabular}{|c|c|c|c|}
        \hline
        \quad\quad$ h_0 $\quad\quad & \quad\quad$q_0$\quad\quad & \quad\quad$h_l$, $q_l$\quad\quad  & $F_l(a=0.75)$ \\
      \hline
		0.00100&	0.15435&	0.00536&		0.03480 \\
        \hline
		0.00200&	0.17143&	0.00882&	0.05148 \\
        \hline
		0.00300&	0.18316&	0.01181&	0.06448\\
        \hline
		0.00400&	0.19270&	0.01455&		0.07552 \\
		\hline
		0.00500&	0.20081&	0.01712&	0.08526 \\
		\hline	
		0.00619&	0.20923&	0.02001&	0.095653 \\
		\hline
		0.00800&	0.22032&	0.02416&	0.10965  \\
		\hline
		0.01073&	0.23450&	0.03000&	0.12793\\
        \hline
        0.01461&	0.25152&	0.03772&	0.15003\\
		\hline
		0.02130&	0.27564&	0.05001&	0.18147 \\
        \hline
        0.02593&	0.29003&	0.05799&		0.20000  \\
        \hline
        0.03155&	0.30585&	0.06727&	0.22002 \\
		\hline
		0.03961&	0.32628&	0.08000&		0.24525   \\
        \hline 
        0.04486&	0.33853&	0.08799&		0.26001 \\
		\hline
		0.05298&	0.35625&	0.10000&	0.28079   \\
        \hline
        0.06129&	0.37318&	0.11192&	0.30002\\
		\hline
		0.07431&	0.39784&	0.13000&	0.32689  \\
		\hline
		0.08922&	0.42401&	0.15000&	0.35394  \\
        \hline
        0.09902&	0.44028&	0.16283&	0.37001 \\
		\hline
		0.11241&	0.46156&	0.18000&	0.39019  \\
		\hline
		0.12833&	0.48575&	0.20000&	0.41196  \\
        \hline 
        0.14049&	0.50355&	0.21500&	0.42723\\
		\hline
		0.15281&	0.52109&	0.23001&	0.44168  \\
		\hline
		0.16946&	0.54412&	0.25000&	0.45979  \\
        \hline
        0.18211&	0.56117&	0.26501&	0.47258 \\
		\hline
		0.19487&	0.57805&	0.28000&	0.48476   \\
		\hline
		0.21207&	0.60035&	0.30000&	0.50012  \\
        \hline
        0.23382&	0.62792&	0.32500&	0.51804 \\
		\hline
		0.25581&	0.65520&	0.35000&	0.53468  \\
        \hline 
        0.27802&	0.68226&	0.37500&	0.55018\\
		\hline
		0.30042&	0.70912&	0.40000&	0.56465  \\
        \hline 
        0.32300&	0.73583&	0.42500&	0.57819 \\
		\hline
		0.34573&	0.76239&	0.45000&	0.59089 \\
        \hline
\end{tabular}
\end{subtable}
\begin{subtable}{0.45\linewidth}
\begin{tabular}{|c|c|c|c|}
\hline
        \quad\quad$h_0$\quad\quad & \quad\quad$q_0$\quad\quad & \quad\quad$h_l$, $q_l$\quad\quad  & $F_l(a=0.75)$ \\
\hline
	0.36860&	0.78882&	0.47500&	0.60283  \\
        \hline
 0.39156&	0.81530&	0.50000&	0.61413 \\
        \hline
0.43797&	0.86735&	0.55000&	0.63483 \\
        \hline
	0.48487&	0.91878&	0.60000&	0.65378 \\
		\hline
0.53243&	0.96876&	0.65000&	0.67148 \\
		\hline	
0.58049&	1.01791&	0.70000&		0.68823 \\
		\hline
0.62913&	1.06588&	0.75000&	0.70420  \\
		\hline
0.6783&	1.11280&	0.80000&	0.71946 \\
        \hline
0.72796&	1.15891&	0.85000&	0.73400 \\
		\hline
 0.77799&	1.20443&	0.90000&	0.74777 \\
        \hline
        0.82828&	1.24972&	0.95000&	0.76068 \\
        \hline
0.87826&	1.29765&	1.00031&	0.77275 \\
		\hline
		 0.97982&	1.38640&	1.10000&	0.79334  \\
        \hline 
 1.08080&	1.47957&	1.20007&	0.81057 \\
		\hline
	 1.18130&	1.57466&	1.30008&		0.82515  \\
        \hline
1.28170&	1.67057&	1.40006&	0.83763 \\
		\hline
 1.38200&	1.76709&	1.50000&	0.84845  \\
		\hline
 1.48240&	1.86424&	1.60009&	0.85793  \\
        \hline
        1.58260&	1.96163&	1.70002&	0.86628 \\
		\hline
 1.68290&	2.05948&	1.80008&	0.87371  \\
		\hline
 1.78310&	2.15752&	1.90007&	0.88036  \\
        \hline 
1.88330&	2.25580&	2.00008&	0.88634 \\
		\hline
   2.38400&	2.74942&	2.50007&	0.90907  \\
		\hline
	 2.88440&	3.24529&	3.00000&	0.92423 \\
        \hline
3.38480&	3.74253&	3.50007&	0.93505 \\
		\hline
 3.88500&	4.24041&	4.00003&	0.94317   \\
		\hline
	 4.38520&	4.73883&	4.50004&	0.94948 \\
        \hline
4.88540&	5.23763&	5.00009&	0.95454 \\
		\hline
	 5.88560&	6.23574&	6.00007&	0.96211 \\
        \hline 
6.88570&	7.23437&	7.00001&	0.96753 \\
		\hline
   7.88580&	8.23338&	8.00000&	0.97158  \\
        \hline 
 8.88590&	9.23263&	9.00001&		0.97474 \\
		\hline
	 9.88600&	10.23206&	10.00003&	0.97727 \\
        \hline
\end{tabular}
\end{subtable}
\end{table}
\begin{figure}
     \begin{subfigure}[!h]{0.49\columnwidth}
         \includegraphics[width=8.4cm]{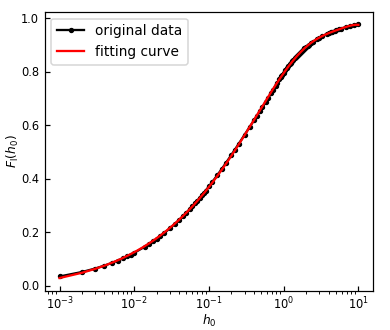}
         \caption{\label{fig:figA1} Fitting equation \ref{eq:eq_A5}}
     \end{subfigure}
     \begin{subfigure}[!h]{0.49\columnwidth}
         \includegraphics[width=8.4cm]{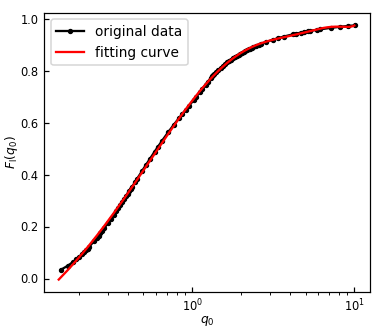}
         \subcaption{\label{fig:figA2} Fitting equation \ref{eq:eq_A6}}
     \end{subfigure}
        \caption{Plots of the fitting Eqs. (\ref{eq:eq_A5} und \ref{eq:eq_A6})}
        \label{fig:fitting_plots}
\end{figure}

\end{widetext}

\clearpage

\nocite{*}

\bibliography{apssamp}

\begin{thebibliography}{26}%
\makeatletter
\providecommand \@ifxundefined [1]{%
 \@ifx{#1\undefined}
}%
\providecommand \@ifnum [1]{%
 \ifnum #1\expandafter \@firstoftwo
 \else \expandafter \@secondoftwo
 \fi
}%
\providecommand \@ifx [1]{%
 \ifx #1\expandafter \@firstoftwo
 \else \expandafter \@secondoftwo
 \fi
}%
\providecommand \natexlab [1]{#1}%
\providecommand \enquote  [1]{``#1''}%
\providecommand \bibnamefont  [1]{#1}%
\providecommand \bibfnamefont [1]{#1}%
\providecommand \citenamefont [1]{#1}%
\providecommand \href@noop [0]{\@secondoftwo}%
\providecommand \href [0]{\begingroup \@sanitize@url \@href}%
\providecommand \@href[1]{\@@startlink{#1}\@@href}%
\providecommand \@@href[1]{\endgroup#1\@@endlink}%
\providecommand \@sanitize@url [0]{\catcode `\\12\catcode `\$12\catcode `\&12\catcode `\#12\catcode `\^12\catcode `\_12\catcode `\%12\relax}%
\providecommand \@@startlink[1]{}%
\providecommand \@@endlink[0]{}%
\providecommand \url  [0]{\begingroup\@sanitize@url \@url }%
\providecommand \@url [1]{\endgroup\@href {#1}{\urlprefix }}%
\providecommand \urlprefix  [0]{URL }%
\providecommand \Eprint [0]{\href }%
\providecommand \doibase [0]{https://doi.org/}%
\providecommand \selectlanguage [0]{\@gobble}%
\providecommand \bibinfo  [0]{\@secondoftwo}%
\providecommand \bibfield  [0]{\@secondoftwo}%
\providecommand \translation [1]{[#1]}%
\providecommand \BibitemOpen [0]{}%
\providecommand \bibitemStop [0]{}%
\providecommand \bibitemNoStop [0]{.\EOS\space}%
\providecommand \EOS [0]{\spacefactor3000\relax}%
\providecommand \BibitemShut  [1]{\csname bibitem#1\endcsname}%
\let\auto@bib@innerbib\@empty
\bibitem [{\citenamefont {Pippard}(1947)}]{pippard1954anomalous}%
  \BibitemOpen
  \bibfield  {author} {\bibinfo {author} {\bibfnamefont {A.~B.}\ \bibnamefont {Pippard}},\ }\bibfield  {title} {\bibinfo {title} {The anomalous skin effect in normal metals},\ }\href@noop {} {\bibfield  {journal} {\bibinfo  {journal} {Proceedings of the Royal Society of London. Series A. Mathematical and Physical Sciences}\ }\textbf {\bibinfo {volume} {191}} (\bibinfo {year} {1947})}\BibitemShut {NoStop}%
\bibitem [{\citenamefont {Reuter}\ and\ \citenamefont {Sondheimer}(1948)}]{reuter1948theory}%
  \BibitemOpen
  \bibfield  {author} {\bibinfo {author} {\bibfnamefont {G.}~\bibnamefont {Reuter}}\ and\ \bibinfo {author} {\bibfnamefont {E.}~\bibnamefont {Sondheimer}},\ }\bibfield  {title} {\bibinfo {title} {The theory of the anomalous skin effect in metals},\ }\href@noop {} {\bibfield  {journal} {\bibinfo  {journal} {Proceedings of the Royal Society of London. Series A. Mathematical and Physical Sciences}\ }\textbf {\bibinfo {volume} {195}},\ \bibinfo {pages} {336} (\bibinfo {year} {1948})}\BibitemShut {NoStop}%
\bibitem [{\citenamefont {Chambers}(1952)}]{chambers1952anomalous}%
  \BibitemOpen
  \bibfield  {author} {\bibinfo {author} {\bibfnamefont {R.}~\bibnamefont {Chambers}},\ }\bibfield  {title} {\bibinfo {title} {The anomalous skin effect},\ }\href@noop {} {\bibfield  {journal} {\bibinfo  {journal} {Proceedings of the Royal Society of London. Series A. Mathematical and Physical Sciences}\ }\textbf {\bibinfo {volume} {215}},\ \bibinfo {pages} {481} (\bibinfo {year} {1952})}\BibitemShut {NoStop}%
\bibitem [{\citenamefont {Jacewicz}\ \emph {et~al.}(2020)\citenamefont {Jacewicz}, \citenamefont {Eriksson}, \citenamefont {Ruber}, \citenamefont {Calatroni}, \citenamefont {Profatilova},\ and\ \citenamefont {Wuensch}}]{jacewicz2020temperature}%
  \BibitemOpen
  \bibfield  {author} {\bibinfo {author} {\bibfnamefont {M.}~\bibnamefont {Jacewicz}}, \bibinfo {author} {\bibfnamefont {J.}~\bibnamefont {Eriksson}}, \bibinfo {author} {\bibfnamefont {R.}~\bibnamefont {Ruber}}, \bibinfo {author} {\bibfnamefont {S.}~\bibnamefont {Calatroni}}, \bibinfo {author} {\bibfnamefont {I.}~\bibnamefont {Profatilova}},\ and\ \bibinfo {author} {\bibfnamefont {W.}~\bibnamefont {Wuensch}},\ }\bibfield  {title} {\bibinfo {title} {Temperature-dependent field emission and breakdown measurements using a pulsed high-voltage cryosystem},\ }\href@noop {} {\bibfield  {journal} {\bibinfo  {journal} {Physical Review Applied}\ }\textbf {\bibinfo {volume} {14}},\ \bibinfo {pages} {061002} (\bibinfo {year} {2020})}\BibitemShut {NoStop}%
\bibitem [{\citenamefont {Nordlund}\ and\ \citenamefont {Djurabekova}(2012)}]{nordlund2012defect}%
  \BibitemOpen
  \bibfield  {author} {\bibinfo {author} {\bibfnamefont {K.}~\bibnamefont {Nordlund}}\ and\ \bibinfo {author} {\bibfnamefont {F.}~\bibnamefont {Djurabekova}},\ }\bibfield  {title} {\bibinfo {title} {Defect model for the dependence of breakdown rate on external electric fields},\ }\href@noop {} {\bibfield  {journal} {\bibinfo  {journal} {Physical Review Special Topics-Accelerators and Beams}\ }\textbf {\bibinfo {volume} {15}},\ \bibinfo {pages} {071002} (\bibinfo {year} {2012})}\BibitemShut {NoStop}%
\bibitem [{\citenamefont {Vernieri}\ \emph {et~al.}(2023)\citenamefont {Vernieri}, \citenamefont {Nanni} \emph {et~al.}}]{Vernieri_2023}%
  \BibitemOpen
  \bibfield  {author} {\bibinfo {author} {\bibfnamefont {C.}~\bibnamefont {Vernieri}}, \bibinfo {author} {\bibfnamefont {E.~A.}\ \bibnamefont {Nanni}}, \emph {et~al.},\ }\bibfield  {title} {\bibinfo {title} {A “cool” route to the higgs boson and beyond. the cool copper collider},\ }\href {https://doi.org/doi:10.1088/1748-0221/18/07/P07053} {\bibfield  {journal} {\bibinfo  {journal} {Journal of Instrumentation}\ }\textbf {\bibinfo {volume} {18}}\bibinfo  {number} { (07)},\ \bibinfo {pages} {P07053}}\BibitemShut {NoStop}%
\bibitem [{\citenamefont {Nasr}\ \emph {et~al.}(2021)\citenamefont {Nasr}, \citenamefont {Nanni}, \citenamefont {Breidenbach}, \citenamefont {Weathersby}, \citenamefont {Oriunno},\ and\ \citenamefont {Tantawi}}]{nasr2021experimental}%
  \BibitemOpen
\bibfield  {number} {  }\bibfield  {author} {\bibinfo {author} {\bibfnamefont {M.}~\bibnamefont {Nasr}}, \bibinfo {author} {\bibfnamefont {E.}~\bibnamefont {Nanni}}, \bibinfo {author} {\bibfnamefont {M.}~\bibnamefont {Breidenbach}}, \bibinfo {author} {\bibfnamefont {S.}~\bibnamefont {Weathersby}}, \bibinfo {author} {\bibfnamefont {M.}~\bibnamefont {Oriunno}},\ and\ \bibinfo {author} {\bibfnamefont {S.}~\bibnamefont {Tantawi}},\ }\bibfield  {title} {\bibinfo {title} {Experimental demonstration of particle acceleration with normal conducting accelerating structure at cryogenic temperature},\ }\href@noop {} {\bibfield  {journal} {\bibinfo  {journal} {Physical Review Accelerators and Beams}\ }\textbf {\bibinfo {volume} {24}},\ \bibinfo {pages} {093201} (\bibinfo {year} {2021})}\BibitemShut {NoStop}%
\bibitem [{\citenamefont {Braun}\ \emph {et~al.}(2003)\citenamefont {Braun}, \citenamefont {D{\"o}bert}, \citenamefont {Wilson},\ and\ \citenamefont {Wuensch}}]{braun2003frequency}%
  \BibitemOpen
  \bibfield  {author} {\bibinfo {author} {\bibfnamefont {H.~H.}\ \bibnamefont {Braun}}, \bibinfo {author} {\bibfnamefont {S.}~\bibnamefont {D{\"o}bert}}, \bibinfo {author} {\bibfnamefont {I.}~\bibnamefont {Wilson}},\ and\ \bibinfo {author} {\bibfnamefont {W.}~\bibnamefont {Wuensch}},\ }\bibfield  {title} {\bibinfo {title} {Frequency and temperature dependence of electrical breakdown at 21, 30, and 39 \text{GHz}},\ }\href@noop {} {\bibfield  {journal} {\bibinfo  {journal} {Physical review letters}\ }\textbf {\bibinfo {volume} {90}},\ \bibinfo {pages} {224801} (\bibinfo {year} {2003})}\BibitemShut {NoStop}%
\bibitem [{\citenamefont {Cahill}\ \emph {et~al.}(2016)\citenamefont {Cahill} \emph {et~al.}}]{CahillIPAC2016-MOPMW038}%
  \BibitemOpen
  \bibfield  {author} {\bibinfo {author} {\bibfnamefont {A.}~\bibnamefont {Cahill}} \emph {et~al.},\ }\bibfield  {title} {\bibinfo {title} {{M}easurements of copper \text{RF} surface resistance at cryogenic temperatures for applications to {X}-band and {S}-band accelerators},\ }\bibfield  {journal} {\bibinfo  {journal} {Proceedings of the the Seventh International Particle Accelerator Conference, Busan, Korea,}\ }\href {https://doi.org/doi:10.18429/JACoW-IPAC2016-MOPMW038} {doi:10.18429/JACoW-IPAC2016-MOPMW038} (\bibinfo {year} {2016})\BibitemShut {NoStop}%
\bibitem [{\citenamefont {Wang}(2023)}]{thesis2023wang}%
  \BibitemOpen
  \bibfield  {author} {\bibinfo {author} {\bibfnamefont {H.}~\bibnamefont {Wang}},\ }\emph {\bibinfo {title} {Operation of copper cavities at cryogenic temperatures}},\ \href@noop {} {Ph.D. thesis},\ \bibinfo  {school} {Frankfurt am Main. Available from: https://nbn-resolving.org/urn:nbn:de:hebis:30:3-792288} (\bibinfo {year} {2023})\BibitemShut {NoStop}%
\bibitem [{\citenamefont {Wang}\ and\ \citenamefont {Ratzinger}(2023)}]{wangipac2023-tupa189}%
  \BibitemOpen
  \bibfield  {author} {\bibinfo {author} {\bibfnamefont {H.}~\bibnamefont {Wang}}\ and\ \bibinfo {author} {\bibfnamefont {U.}~\bibnamefont {Ratzinger}},\ }\bibfield  {title} {\bibinfo {title} {Operation of copper cavities at cryogenic temperatures},\ }\bibfield  {journal} {\bibinfo  {journal} {Proc. 14th International Particle Accelerator Conference, Venice, Italy,}\ }\href {https://doi.org/doi:10.18429/JACoW-IPAC2023-TUPA189} {doi:10.18429/JACoW-IPAC2023-TUPA189} (\bibinfo {year} {2023})\BibitemShut {NoStop}%
\bibitem [{\citenamefont {Caspers}\ \emph {et~al.}(1997)\citenamefont {Caspers}, \citenamefont {Morvillo},\ and\ \citenamefont {Ruggiero}}]{caspers1997surface}%
  \BibitemOpen
  \bibfield  {author} {\bibinfo {author} {\bibfnamefont {F.}~\bibnamefont {Caspers}}, \bibinfo {author} {\bibfnamefont {M.}~\bibnamefont {Morvillo}},\ and\ \bibinfo {author} {\bibfnamefont {F.}~\bibnamefont {Ruggiero}},\ }\bibfield  {title} {\bibinfo {title} {Surface resistance measurements for the \text{LHC} beam screen},\ }in\ \href@noop {} {\emph {\bibinfo {booktitle} {Proceedings of the 1997 Particle Accelerator Conference (Cat. No. 97CH36167)}}},\ Vol.~\bibinfo {volume} {1}\ (\bibinfo {organization} {IEEE},\ \bibinfo {year} {1997})\ pp.\ \bibinfo {pages} {75--77}\BibitemShut {NoStop}%
\bibitem [{\citenamefont {Weingarten}(1990)}]{weingarten}%
  \BibitemOpen
  \bibfield  {author} {\bibinfo {author} {\bibfnamefont {W.}~\bibnamefont {Weingarten}},\ }\bibfield  {title} {\bibinfo {title} {Radio-frequency superconductivity applied to large electron accelerators},\ }\href@noop {} {\bibfield  {journal} {\bibinfo  {journal} {Particle World}\ }\textbf {\bibinfo {volume} {Vol l. No. 4, p.93-103}} (\bibinfo {year} {1990})}\BibitemShut {NoStop}%
\bibitem [{\citenamefont {Liew}\ \emph {et~al.}(2014)\citenamefont {Liew}, \citenamefont {Malaysia} \emph {et~al.}}]{liew2014signal}%
  \BibitemOpen
  \bibfield  {author} {\bibinfo {author} {\bibfnamefont {E.}~\bibnamefont {Liew}}, \bibinfo {author} {\bibfnamefont {M.~C.~F.}\ \bibnamefont {Malaysia}}, \emph {et~al.},\ }\bibfield  {title} {\bibinfo {title} {Signal transmission loss due to copper surface roughness in high-frequency region},\ }in\ \href@noop {} {\emph {\bibinfo {booktitle} {Proc. IPC APEX EXPO}}}\ (\bibinfo {year} {2014})\BibitemShut {NoStop}%
\bibitem [{\citenamefont {Ratzinger}\ \emph {et~al.}(2019)\citenamefont {Ratzinger}, \citenamefont {H{\"a}hnel}, \citenamefont {Tiede}, \citenamefont {Kaiser},\ and\ \citenamefont {Almomani}}]{ratzinger2019combined}%
  \BibitemOpen
  \bibfield  {author} {\bibinfo {author} {\bibfnamefont {U.}~\bibnamefont {Ratzinger}}, \bibinfo {author} {\bibfnamefont {H.}~\bibnamefont {H{\"a}hnel}}, \bibinfo {author} {\bibfnamefont {R.}~\bibnamefont {Tiede}}, \bibinfo {author} {\bibfnamefont {J.}~\bibnamefont {Kaiser}},\ and\ \bibinfo {author} {\bibfnamefont {A.}~\bibnamefont {Almomani}},\ }\bibfield  {title} {\bibinfo {title} {Combined zero degree structure beam dynamics and applications},\ }\href@noop {} {\bibfield  {journal} {\bibinfo  {journal} {Physical Review Accelerators and Beams}\ }\textbf {\bibinfo {volume} {22}},\ \bibinfo {pages} {114801} (\bibinfo {year} {2019})}\BibitemShut {NoStop}%
\bibitem [{\citenamefont {Broere}\ \emph {et~al.}(1998)\citenamefont {Broere}, \citenamefont {Kugler}, \citenamefont {Krietenstein}, \citenamefont {Ratzinger},\ and\ \citenamefont {Vretenar}}]{broere1998high}%
  \BibitemOpen
  \bibfield  {author} {\bibinfo {author} {\bibfnamefont {J.}~\bibnamefont {Broere}}, \bibinfo {author} {\bibfnamefont {H.}~\bibnamefont {Kugler}}, \bibinfo {author} {\bibfnamefont {B.}~\bibnamefont {Krietenstein}}, \bibinfo {author} {\bibfnamefont {U.}~\bibnamefont {Ratzinger}},\ and\ \bibinfo {author} {\bibfnamefont {M.}~\bibnamefont {Vretenar}},\ }\bibfield  {title} {\bibinfo {title} {High power conditioning of the 202 mhz ih tank 2 at the cern linac3},\ }\href {https://accelconf.web.cern.ch/l98/PAPERS/TH4004.PDF} {\bibfield  {journal} {\bibinfo  {journal} {19th International Linear Accelerator Conference, Chicago, IL, USA,}\ } (\bibinfo {year} {1998})}\BibitemShut {NoStop}%
\bibitem [{\citenamefont {H{\"a}hnel}\ \emph {et~al.}(2023)\citenamefont {H{\"a}hnel}, \citenamefont {Ate{\c{s}}}, \citenamefont {Dedi{\'c}},\ and\ \citenamefont {Ratzinger}}]{hahnel2023additive}%
  \BibitemOpen
  \bibfield  {author} {\bibinfo {author} {\bibfnamefont {H.}~\bibnamefont {H{\"a}hnel}}, \bibinfo {author} {\bibfnamefont {A.}~\bibnamefont {Ate{\c{s}}}}, \bibinfo {author} {\bibfnamefont {B.}~\bibnamefont {Dedi{\'c}}},\ and\ \bibinfo {author} {\bibfnamefont {U.}~\bibnamefont {Ratzinger}},\ }\bibfield  {title} {\bibinfo {title} {Additive manufacturing of an ih-type linac structure from stainless steel and pure copper},\ }\href@noop {} {\bibfield  {journal} {\bibinfo  {journal} {Instruments}\ }\textbf {\bibinfo {volume} {7}},\ \bibinfo {pages} {22} (\bibinfo {year} {2023})}\BibitemShut {NoStop}%
\bibitem [{\citenamefont {Fouaidy}\ and\ \citenamefont {Hammoudi}(2006)}]{fouaidy2006rrr}%
  \BibitemOpen
  \bibfield  {author} {\bibinfo {author} {\bibfnamefont {M.}~\bibnamefont {Fouaidy}}\ and\ \bibinfo {author} {\bibfnamefont {N.}~\bibnamefont {Hammoudi}},\ }\bibfield  {title} {\bibinfo {title} {\text{RRR} of copper coating and low temperature electrical resistivity of material for \text{TTF} couplers},\ }\href@noop {} {\bibfield  {journal} {\bibinfo  {journal} {Physica C: Superconductivity}\ }\textbf {\bibinfo {volume} {441}},\ \bibinfo {pages} {137} (\bibinfo {year} {2006})}\BibitemShut {NoStop}%
\bibitem [{\citenamefont {Matula}(1979)}]{matula1979electrical}%
  \BibitemOpen
  \bibfield  {author} {\bibinfo {author} {\bibfnamefont {R.~A.}\ \bibnamefont {Matula}},\ }\bibfield  {title} {\bibinfo {title} {Electrical resistivity of copper, gold, palladium, and silver},\ }\href@noop {} {\bibfield  {journal} {\bibinfo  {journal} {Journal of Physical and Chemical Reference Data}\ }\textbf {\bibinfo {volume} {8}},\ \bibinfo {pages} {1147} (\bibinfo {year} {1979})}\BibitemShut {NoStop}%
\bibitem [{\citenamefont {Jackson}(1999)}]{jackson1999classical}%
  \BibitemOpen
  \bibfield  {author} {\bibinfo {author} {\bibfnamefont {J.~D.}\ \bibnamefont {Jackson}},\ }\href@noop {} {\emph {\bibinfo {title} {Classical electrodynamics}}}\ (\bibinfo  {publisher} {American Association of Physics Teachers},\ \bibinfo {year} {1999})\BibitemShut {NoStop}%
\bibitem [{\citenamefont {Ashcroft}\ and\ \citenamefont {Mermin}(1988)}]{ashcroft1976solid}%
  \BibitemOpen
  \bibfield  {author} {\bibinfo {author} {\bibfnamefont {N.~W.}\ \bibnamefont {Ashcroft}}\ and\ \bibinfo {author} {\bibfnamefont {N.~D.}\ \bibnamefont {Mermin}},\ }\href@noop {} {\emph {\bibinfo {title} {Solid State Physics}}},\ \bibinfo {edition} {internat.}\ ed.\ (\bibinfo  {publisher} {Saunders College},\ \bibinfo {year} {1988})\BibitemShut {NoStop}%
\bibitem [{\citenamefont {{P}rivate communication}(2023)}]{szcepaniak}%
  \BibitemOpen
  \bibfield  {author} {\bibinfo {author} {\bibnamefont {{P}rivate communication}},\ }\href@noop {} {\emph {\bibinfo {title} {Company Galvano-T, Windeck, Germany}}},\ \bibinfo {type} {Tech. Rep.}\ (\bibinfo {year} {2023})\BibitemShut {NoStop}%
\bibitem [{\citenamefont {Ermakov}(2014)}]{ermakov2014}%
  \BibitemOpen
  \bibfield  {author} {\bibinfo {author} {\bibfnamefont {A.}~\bibnamefont {Ermakov}},\ }\href@noop {} {\emph {\bibinfo {title} {TTF-Coupler - Testing of the Copper Layer - internal report}}},\ \bibinfo {type} {Tech. Rep.}\ (\bibinfo {year} {2014})\BibitemShut {NoStop}%
\bibitem [{\citenamefont {Holman}(2010)}]{holman2010heat}%
  \BibitemOpen
  \bibfield  {author} {\bibinfo {author} {\bibfnamefont {J.~P.}\ \bibnamefont {Holman}},\ }\href@noop {} {\emph {\bibinfo {title} {Heat transfer}}},\ \bibinfo {edition} {10th}\ ed.\ (\bibinfo  {publisher} {McGraw Hill Higher Education},\ \bibinfo {year} {2010})\ p.\ \bibinfo {pages} {139}\BibitemShut {NoStop}%
\bibitem [{\citenamefont {Almomani}\ and\ \citenamefont {Ratzinger}()}]{AlmomaniIPAC2014}%
  \BibitemOpen
  \bibfield  {author} {\bibinfo {author} {\bibfnamefont {A.}~\bibnamefont {Almomani}}\ and\ \bibinfo {author} {\bibfnamefont {U.}~\bibnamefont {Ratzinger}},\ }\bibfield  {title} {\bibinfo {title} {A 325 {MH}z high gradient {CH} – test cavity for $\beta$=0.16},\ }in\ \href {https://doi.org/https://doi.org/10.18429/JACoW-IPAC2014-THPME010} {\emph {\bibinfo {booktitle} {Proc. 5th International Particle Accelerator Conference (IPAC'14), Dresden, Germany, June 15-20, 2014}}},\ \bibinfo {note} {https://doi.org/10.18429/JACoW-IPAC2014-THPME010}\BibitemShut {NoStop}%
\bibitem [{\citenamefont {Dalakov}\ \emph {et~al.}(2020)\citenamefont {Dalakov}, \citenamefont {Neuber}, \citenamefont {Klier},\ and\ \citenamefont {Herzog}}]{dalakov2020innovative}%
  \BibitemOpen
  \bibfield  {author} {\bibinfo {author} {\bibfnamefont {P.}~\bibnamefont {Dalakov}}, \bibinfo {author} {\bibfnamefont {E.}~\bibnamefont {Neuber}}, \bibinfo {author} {\bibfnamefont {J.}~\bibnamefont {Klier}},\ and\ \bibinfo {author} {\bibfnamefont {R.}~\bibnamefont {Herzog}},\ }\bibfield  {title} {\bibinfo {title} {Innovative neon refrigeration unit operating down to 30 k},\ }in\ \href@noop {} {\emph {\bibinfo {booktitle} {MATEC Web of Conferences}}},\ Vol.\ \bibinfo {volume} {324}\ (\bibinfo {organization} {EDP Sciences},\ \bibinfo {year} {2020})\ p.\ \bibinfo {pages} {01003}\BibitemShut {NoStop}%
\end{thebibliography}%

\end{document}